\documentclass[12pt,epsf]{article}
\pdfoutput=1
\usepackage[pdftex]{graphicx}
\usepackage{amsmath,amssymb}
\usepackage{bm}
\usepackage{amsfonts}
\usepackage{mathrsfs}
\usepackage{ascmac}
\usepackage{braket}
\graphicspath{{./fig/}}
\usepackage{comment}
\usepackage{cite}

\newcommand{\beq}{\begin{equation}}
\newcommand{\eeq}{\end{equation}}
\newcommand{\bea}{\begin{eqnarray}}
\newcommand{\eea}{\end{eqnarray}}

\newcommand{\ba}{\begin{aligned}}
\newcommand{\ea}{\end{aligned}}

\def\pe2{p_E^2}

\begin{document}
\setlength{\baselineskip}{0.7cm}
\begin{titlepage} 
\begin{flushright}
OCU-PHYS 528  \\
NITEP 86
\end{flushright}
\vspace*{10mm}%
\begin{center}{\LARGE\bf
Non-Gaussianity from $X, Y$ gauge bosons \\
\vspace*{3mm}
in Cosmological Collider Physics
}
\end{center}
\vspace*{10mm}
\begin{center}
{\Large Nobuhito Maru}$^{a,b}$ and 
{\Large Akira Okawa}$^{a}$ 
\end{center}
\vspace*{0.2cm}
\begin{center}
${}^{a}${\it 
Department of Mathematics and Physics, Osaka City University, \\ 
Osaka 558-8585, Japan}
\\
${}^{b}${\it Nambu Yoichiro Institute of Theoretical and Experimental Physics (NITEP), \\
Osaka City University, 
Osaka 558-8585, Japan} 
\end{center}
\vspace*{1cm}

\begin{abstract} 
Heavy fields of Hubble scale order present during inflation 
 contribute to the non-Gaussian signature for the three-point function of the inflaton. 
Taking into account that Hubble scale is around the scale of grand unified theory (GUT), 
 this opens a possibility that the GUT scale signatures, which are very hard to be discovered at collider, 
 might be detectable by using information from the precise observations of cosmic microwave background. 
We discuss a detactability of the $X, Y$ gauge boson present in any GUT 
 in a framework of cosmological collider physics. 
Calculating one-loop contributions of $X, Y$ gauge bosons to the inflaton three-point functions, 
 we find a remarkable result that 
 one-loop diagram with interactions originated from the mass terms of $X, Y$ gauge bosons 
 provides an enhancement factor expressed by the ratio 
 between the $X, Y$ gauge boson mass and Hubble scale as $(m_X/H)^4$. 
In an estimation of the non-Gaussianity, 
this factor is crucial and its impact on the detactability of $X, Y$ gauge bosons is discussed.    
\end{abstract}
\end{titlepage}

\section{Introduction} 
The standard model (SM) of particle physics has described many physical phenomena and has gained trust. 
However, some observations cannot be explained by the SM, 
 such as neutrino oscillation and dark matter problems. 
For this reason, several theories beyond the standard model have been studied, 
 for example, a Grand Unified Theory (GUT) that unifies the electromagnetic, strong and weak forces. 
Any GUT includes $X, Y$ gauge bosons that do not exist in the SM, 
 which are the massive gauge bosons in a spontaneous breaking of the GUT gauge group to the SM ones. 
Typically the prediction of GUT is a proton decay by mediating these particles, 
 but has not yet been observed in Super-Kamiokande experiment. 
According to the gauge coupling evolution via renormalization group equation, 
 the GUT scale is predicted to be around $10^{15}$ GeV, 
 and it is very difficult to confirm such a high energy theory signature at collider experiments. 

On the other hand, it has been pointed out that the development of cosmological precision observation
 in recent years has the potential to verify high-energy physics. 
Precise observation of cosmic microwave background radiation (CMB) by Planck satellite 
 has confirmed the inflation mechanism that solves the flatness problem, horizon problem, and monopole problem. 
It has been estimated from the observations that inflation occurs at an energy scale (inflation scale) 
 of about $ 10 ^ {14} $ GeV by a scalar field called inflaton, 
 which is very close to the energy scale of the GUT. 
 For this reason, an approach called Cosmological Collider Physics in elementary particle physics 
  has been actively studied in recent years \cite{ChenWang1, ChenWang2, BaumannGreen, ABG, 
  SFCS, NVBB, ChenWang3, NYY, GPS, Emami, KehagiasRiotto1, LWZ, NAHM, DFK, SCD, CNW, DNS, 
  BBDS, FMSS, LBP, DGS, MMMC, CWX1, CWX2, KehagiasRiotto2, AMRW1, TWZ, IWWZ, AMRW2, 
  KumarSundrum1, Riquelme, FKR, SaitoKubota, CPS, WWYZ, CWX3, DFT, BCKS,CDWZ, NAHBLP, 
  KumarSundrum2, GHJT, Wu, ADLFKR, NSWZ, MR, KNTZ, LWX, HookHaungRacco1, HookHuangRacco2, 
  KumarSundrum3, WangXianyu1, WangZhu, LLWZ, WangXianyu2, BCS}. 
This approach uses the effective field theory of inflation, 
 which integrates inflation scale heavy particles other than inflaton and graviton. 
Heavy particles of the inflation scale order contribute to a three-point correlation function 
 that describes the interaction between inflaton and graviton. 
If these three-point functions give the non-Gaussian property, 
 which is the deviation of the statistical distribution of the curvature fluctuation from the Gaussian distribution, 
 information on the particles that existing during the inflation period can be obtained 
 by comparing it with observation data such as Planck satellites. 
Therefore, this approach can regard the universe as a high-energy accelerator 
 and has the advantage that it can extract information on physics 
 at energy scales which cannot be reached by terrestrial accelerators. 

The three-point function of the model with only inflaton and graviton was calculated by Maldacena \cite{Maldacena}. 
The effective field theory of inflation was proposed by Chueng et al. \cite{Cheung}, 
 and models with other particles also began to be considered. 
However, the contribution from the $X, Y$ gauge bosons to the three-point function, 
 whose existence is predicted by a GUT with an energy scale very close to the inflation scale, 
 has not yet been calculated. 
Therefore, 
 we introduce $X, Y$ gauge bosons into our theory in addition to the inflaton and the graviton, 
 and calculate three-point functions in a framework of quantum field theory. 
By comparing the non-Gaussianity of these three-point functions obtained in this paper 
 with observation data such as Planck satellites, 
 we will explore an evidence for heavy particles existing in the Grand Unified Theory 
 and discuss its detectability. 

The organization of this paper is as follows. 
In the next section, our setup is introduced, 
 which includes a brief review of in-in formalism, 
 a derivation of the propagators of $X, Y$  gauge bosons 
 and its interaction terms to the graviton.  
In section 3, the non-Gaussianity is estimated 
 by calculating one-loop contributions from $X, Y$ gauge bosons to the graviton three-point function 
 and converting the results to that of the inflaton three-point function. 
Our conclusion is given in a section 4. 
The detail calculations of one-loop Feynman diagrams relevant to the graviton three-point function 
 are explained in Appendix.   

\section{Set up}

\subsection{Action}
We consider a fluctuating spacetime in ADM form 
 in the Friedmann-Lemaitre-Robertson-Walker (FLRW) metric of curvature $K=0$ 
\begin{equation}
ds^{2} = -N^{2}dt^{2} + \tilde{h}_{ij} \left( dx^{i} +N^{i}dt  \right) \left( dx^{j} +N^{j}dt  \right), 
\end{equation}
where $\tilde{h}_{ij}$ is a spatial components of the metric, 
 $N$ is a lapse function and $N^{i}$ is a shift function.  
The quantity with tilde represents that in the comoving gauge. 
From the viewpoint of the effective field theory \cite{Cheung, Leonardo}, 
 the Einstein-Hilbert action and the inflaton action can be written as
\begin{equation}
\begin{aligned} 
S_{\mathrm{grav}} = \int d^{4} x \sqrt{-g} 
 &\left[\frac{1}{2} M_{\mathrm{Pl}}^{2} R - M_{\mathrm{Pl}}^{2} \left(3 H^{2}(\tilde{t}+\pi)
 +\dot{H}( \tilde{t}+\pi)\right) \right. \\
 & \left. +M_{\mathrm{PI}}^{2} \dot{H}( \tilde{t}+\pi)\left(\partial_{\mu}( \tilde{t}+\pi) \partial_{\nu}
 ( \tilde{t}+\pi) g^{\mu \nu}\right) \right. \\
 & +\frac{M_{2}( \tilde{t}+\pi)^{4}}{2 !}\left(\partial_{\mu}( \tilde{t}+\pi) \partial_{\nu}( \tilde{t}+\pi) 
 g^{\mu \nu}+1\right)^{2}\\
 &+ \left.\frac{M_{3}( \tilde{t}+\pi)^{4}}{3 !}\left(\partial_{\mu}( \tilde{t}+\pi) \partial_{\nu}( \tilde{t}+\pi) 
 g^{\mu \nu}+1\right)^{3}+\ldots
 \right], 
 \end{aligned}
\end{equation}
where $g$ is the determinant of the metric $g_{\mu\nu}$, 
 $M_{\mathrm{Pl}}$ is the Planck mass, 
 $R$ is the Ricci curvature in four dimensions, $H$ is the Hubble parameter, 
 $\pi$ is the inflaton as a Nambu-Goldstone boson of time translation, 
 and $M_{2, 3}$ are the coefficients of the high-dimensional operators. 
Of these, the four-dimensional Ricci curvature $R$ can be decomposed into 
 a three-dimensional Ricci curvature $R^{(3)}$ and an external curvature part $E_{ij}$. 
That is, using the external curvature part
\begin{equation}
E_{i j} \equiv \frac{1}{2}\left(\dot{\tilde{h}}_{i j}-\nabla_{i} N_{j}-\nabla_{j} N_{i}\right)
\end{equation}
with a covariant derivative 
\begin{equation}
\nabla_{\mu} A_{\nu} = \partial_{\mu} A_{\nu} - {\Gamma^{\alpha}}_{\mu\nu}A_{\alpha}
\end{equation}
for gravity ($A_{\mu}$ is a covariant component of a 4-dimensional vector), 
the four-dimensional Ricci curvature can be represented as
\begin{equation}
\sqrt{-g} R = \sqrt{\tilde{h}} \left[ N R^{(3)} + N^{-1} \left( E_{ij}E^{ij} - E^{2} \right)    \right],
\end{equation}
where $\tilde{h}$ is the determinant of the spatial metric $\tilde{h}_{ij}$ 
 and $E$ is defined to be the trace of $E_{ij}$, $E \equiv {E^{i}}_{i}$ taken by the spatial metric $\tilde{h}_{ij}$. 
The specific formula for these quantities are \cite{Ryo Saito}
\begin{eqnarray}
R^{(3)} &=& -\frac{1}{4a^{2}} (\tilde{t}) \partial_{l} \gamma_{i j} \partial_{l} \gamma_{i j}
 +\mathcal{O}\left(\gamma^{3}\right),  \\
E_{i j} E^{i j}-E^{2} &=& -6  {H^{2}+4 H \partial_{i} N^{i}} 
 +\frac{1}{2} (\partial_{i} N^{j})\left(\partial_{i} N^{j}+\partial_{j} N^{i}\right)
 -\left(\partial_{i} N^{i}\right)^{2} 
 \nonumber \\
 &&-\frac{1}{2}\left(\partial_{i} N^{j}+\partial_{j} N^{i}\right) \dot{\gamma}_{i j} 
+\frac{1}{4} \dot{\gamma}_{i j} \dot{\gamma}_{i j}-\frac{1}{2} \dot{\gamma}_{i j} (\partial_{l} \gamma_{i j}) N^{l} 
+(\partial_{l} \gamma_{i j} )N^{l} \partial_{i} N^{j}
\nonumber \\
&& +\mathcal{O} \left(\gamma^{3}\right),
\end{eqnarray}
where the spatial metric $\tilde{h}_{ij}$ has been expanded as 
\begin{equation}
\tilde{h}_{ij} = \delta_{ij} + \tilde{\gamma}_{ij} + \frac{1}{2}\tilde{\gamma}_{ik}\tilde{\gamma}_{kj}.
\label{htilde}
\end{equation}
Since the indices of the perturbations are raised or lowered by $\delta_{ij}$, 
 we do not care about the raising and lowering the indices. 
We also use that the inverse of the metric is given by
\begin{equation}
\tilde{g}^{\mu \nu} = \frac{1}{N^{2}} 
\left(
\begin{array}{cc}
{-1} & {N^{i}} \\ 
{N^{j}} & {N^{2} \tilde{h}^{i j}-N^{i} N^{j}}
\end{array}
\right).
\label{inversemetric}
\end{equation}
Since we consider the energy scale after the GUT gauge symmetry is broken, 
 the action of the $X, Y$ gauge bosons in GUT\footnote{We implicitly 
 consider an $SU(5)$ GUT case, but our results are not essentially changed 
 even if the theory in larger GUT gauge group is considered.} 
 (collectively denoted by $X^a_\mu$) is given by
\begin{eqnarray}
S_{\mathrm{gauge}} &=& \int d^{4}x \sqrt{-g}
\left(
-\frac{1}{4} g^{\mu\rho} g^{\nu\sigma} F^{a}_{\mu\nu} F^{a}_{\rho\sigma} 
- m^{2}_{X} g^{\mu\nu} X^{a}_{\mu} X^{*}_{a\nu} 
\right) 
\nonumber \\
&\supset& \int d^{4}x \sqrt{-g}\left[ -g^{\mu\rho}g^{\nu\sigma} 
\left\{ 
(\partial_{\mu}X_{a\nu})(\partial_{\rho}X^{*}_{a\sigma}) 
- (\partial_{\mu}X_{a\nu})(\partial_{\sigma}X^{*}_{a\rho}) 
\right\} \right. 
\nonumber \\
&{}& \left. 
-  m^{2}_{X}g^{\mu\nu}X^{a}_{\mu}X^{*}_{a\nu} + \cdots 
\right],
\end{eqnarray}
where the field strength of $X,Y$ gauge bosons is 
\begin{align}
F_{\mu\nu} = \partial_\mu X_\nu - \partial_\nu X_\mu - ig_{\textrm{GUT}} [X_\mu, X_\nu] 
\end{align}
with a GUT gauge coupling constant $g_{\textrm{GUT}}$. 
$m_{X}$ is the mass of the $X, Y$ gauge bosons 
 and $a$ is the index for the gauge group. 
$\cdots$ represents interaction terms form the commutator in the field strength $F_{\mu\nu}$, 
 which are neglected here because they are sub-leading effects in our analysis later. 
Since the FLRW background space-time is conformal invariant, 
 we introduce a gauge fixed term
\begin{eqnarray}
S_{\mathrm{GF}} &=& -\frac{1}{\xi} \int d^{4}x \sqrt{-g} 
\left( 
g^{\mu\nu}\nabla_{\mu}X^{a}_{\nu} + g^{\mu\nu}{\Gamma^{\alpha}}_{\mu\nu}X^{a}_{\alpha} 
\right)^{2} \nonumber \\
&=&  -\frac{1}{\xi} \int d^{4}x \sqrt{-g} \left( g^{\mu\nu}\partial_{\mu}X^{a}_{\nu} \right)^{2},
\end{eqnarray}
which preserves conformal invariance. 
$\xi$ is a gauge parameter, which will be fixed to $\xi=1$. 

From equations (\ref{htilde}) and (\ref{inversemetric}), 
 the action of the $X, Y$ gauge bosons with quadratic terms is eventually
\begin{eqnarray}
S_{G} &=&S_{\mathrm{gauge}}  +S_{\mathrm{GF}}   
\nonumber \\
&\supset& \int d^{4}x \sqrt{-g}
\left[ 
-g^{\mu\rho}g^{\nu\sigma} 
\left\{  
(\partial_{\mu}X_{a\nu})(\partial_{\rho}X^{*}_{a\sigma}) 
- (\partial_{\mu}X_{a\nu})(\partial_{\sigma}X^{*}_{a\rho})  
\right\} -  m^{2}_{X}g^{\mu\nu}X^{a}_{\mu}X^{*}_{a\nu}   
\right] \nonumber \\
&{}& -\frac{1}{\xi} \int d^{4}x \sqrt{-g} 
\left( 
g^{\mu\nu}\partial_{\mu}X^{a}_{\nu} 
\right) 
\left(
g^{\rho\sigma}\partial_{\rho}X^{a*}_{\sigma} 
\right).
\label{XYaction}
\end{eqnarray}

\subsection{Propagators of $X, Y$ gauge bosons}
In this subsection, we derive the propagators of the $X, Y$ gauge bosons in inflationary spacetime. 
In order to derive them, it is necessary to introduce briefly a concept of in-in formalism \cite{in-in}. 
For details, please refer to \cite{SK}. 

\begin{figure}[htbp]
 \begin{center}
  \includegraphics[width=100mm]{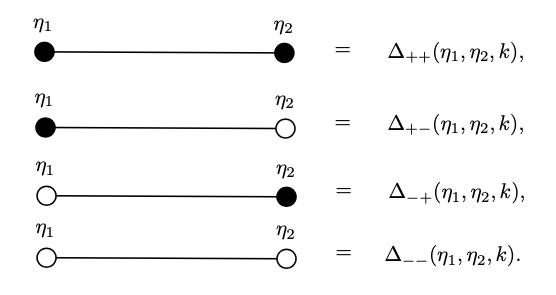}
 \end{center}
 \vspace*{-5mm}
 \caption{Graphical representation for an internal line in in-in formalism.}
\label{Fig1}
 \end{figure}

\begin{figure}[htbp]
 \begin{center}
  \includegraphics[width=100mm]{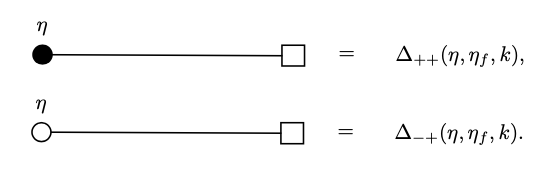}
 \end{center}
 \vspace*{-5mm}
 \caption{Graphical representation for an external line in in-in formalism.}
\label{Fig2}
 \end{figure}

Starting from the Heisenberg picture, 
 let us consider the path integral formalism of the expectation value $\langle Q(\eta)\rangle$ for the operator
\begin{equation}
Q(\eta) \equiv \varphi^{A_{1}}\left(\eta, \bm{x}_{1}\right) \cdots \varphi^{A_{N}}\left(\eta, \bm{x}_{N}\right).
\end{equation}
$\eta$ means a conformal time defined as $d^2 \eta = a^2(t) dt^2$. 
In the Heisenberg picture, the initial state is independent of time, 
 and the expectation value $\langle Q(\eta)\rangle$ is given by
\begin{equation}
\langle Q\rangle=\langle\Omega|Q(\eta)| \Omega\rangle.
\end{equation}
The state $\ket{\Omega}$ is determined at some initial time slice $\eta=\eta_{0}$, 
 and the vacuum state at $\eta=\eta_{0}$ is usually adopted. 
It is natural to adopt the state $\ket{\Omega}$ as the in-state, 
 but since inflationary spacetime is in non-equilibrium, it is nontrivial what the out-state should be. 
To solve this problem, we choose a time slice $\Sigma_{f}$ at an arbitrary time $\eta_{f} \geq \eta$ 
 and insert a complete system 
\begin{equation}
1=\int \prod_{\bm{x}} dO_{\alpha}(\eta_{f}, \bm{x}) 
\left|O_{\alpha}(\eta_{f}, \bm{x}) \right\rangle\left\langle O_{\alpha}(\eta_{f}, \bm{x})\right|
\end{equation}
of local operators $O_{\alpha}\left(\eta, \bm{x}_{i}\right)$ consisting of field operators in the Lagrangian, 
\begin{equation}
\langle Q\rangle=\int \prod_{\bm{x}} dO_{\alpha}(\eta_{f}, \bm{x}) 
\left\langle\Omega \left|O_{\alpha}(\eta_{f}, \bm{x}) \right\rangle 
\left\langle O_{\alpha}(\eta_{f}, \bm{x})\right|  \left. \varphi^{A_{1}}\left(\eta, \bm{x}_{1}\right) 
\cdots \varphi^{A_{N}}\left(\eta, \bm{x}_{N}\right)\right| \Omega\right\rangle.
\end{equation}
Furthermore, we introduce infinitely many slices 
 between the initial slice $\Sigma_{0}$ and the final slice $\Sigma_{f}$, 
 and insert a complete system
\begin{equation}
1=\int \prod_{\bm{x}} d\varphi(\eta_{i}, \bm{x}) 
\left|\varphi(\eta_{i}, \bm{x}) \right\rangle\left\langle \varphi(\eta_{i}, \bm{x})\right|
\end{equation}
of field operators $\varphi^{A}$ and a complete system
\begin{equation}
1=\int \prod_{\bm{x}} d\pi(\eta_{i}, \bm{x}) 
\left|\pi(\eta_{i}, \bm{x}) \right\rangle\left\langle \pi(\eta_{i}, \bm{x})\right|
\end{equation}
of their conjugate operators $\pi_{A}$ in each slice $\Sigma_{i}$. 
If we denote the field on the time-ordered factors $\left\langle O_{\alpha}\left(\eta_{f}\right)|Q(\eta)| 
\Omega\left(\eta_{0}\right)\right\rangle $ by the subscript $+$, we obtain
\begin{eqnarray}
 \left\langle O_{\alpha}\left(\eta_{f}\right)|Q(\eta)| \Omega\left(\eta_{0}\right)\right\rangle  
 &=& \int \mathcal{D} \varphi_{+} \mathcal{D} \pi_{+} 
 \exp \left[ i \int_{\eta_{0}}^{\eta_{f}} d \eta d^{3} x\,\left(\pi_{+A} \varphi_{+}^{\prime A}
 -\mathscr{H}\left[\pi_{+}, \varphi_{+}\right]\right)\right] 
 \nonumber \\
 &{}&  \times \varphi_{+}^{A_{1}}\left(\eta, \mathrm{x}_{1}\right) \cdots \varphi_{+}^{A_{N}}
 \left(\eta, \mathrm{x}_{N}\right)\left\langle O_{\alpha}(\eta_{0}, \bm{x}) | \varphi_{+}\left(\eta_{f}\right)
 \right\rangle\left\langle\varphi_{+}\left(\eta_{0}\right) | \Omega\right\rangle ,
 \nonumber \\
&{}&
\label{20}
\end{eqnarray}
where the prime of $\varphi_+^{\prime A }$ means a derivative with respect to $\eta$. 
$\mathscr{H}\left[\pi_{+}, \varphi_{+}\right]$ is a Hamiltonian 
 and the integral measures are collectively denoted by
\begin{equation}
\mathcal{D} \varphi \mathcal{D} \pi \equiv \lim _{N \rightarrow \infty} \prod_{n=0}^{N-1} \prod_{\bm{x}} 
\frac{d \varphi\left(\eta_{n}, \bm{x}\right)  d \pi\left(\eta_{n}, \bm{x}\right)   }{2\pi} 
=  \prod_{\eta_{0} \leq \eta \leq \eta_{f} }  \prod_{\bm{x}} 
\frac{d \varphi\left(\eta_{n}, \bm{x}\right)  d \pi\left(\eta_{n}, \bm{x}\right)   }{2\pi}.
\end{equation}
Similarly, if we denote the field for anti-time-ordered factors 
 $\left\langle\Omega\left(\eta_{0}\right) | O_{\alpha}(\eta_{f}, \bm{x})\right\rangle $ 
 with a subscript $-$, we obtain
\begin{eqnarray}
\left\langle\Omega\left(\eta_{0}\right) | O_{\alpha}(\eta_{f}, \bm{x})\right\rangle 
&=& \int \mathcal{D} \varphi_{-} \mathcal{D} \pi_{-} 
\exp \left[- i \int_{\eta_{0}}^{\eta_{f}} d \eta d^{3}x\,\left(\pi_{-A} \varphi_{-}^{\prime A}
-\mathscr{H}\left[\pi_{-}, \varphi_{-}\right]\right)\right] 
\nonumber \\
&{}& \vspace{3cm} \times\left\langle\varphi_{-}
\left(\eta_{f}\right) | O_{\alpha}(\eta_{0}, \bm{x})\right\rangle\left\langle\Omega | 
\varphi_{-}\left(\eta_{0}\right)\right\rangle . 
\label{22}
\end{eqnarray}
Putting the equations (\ref{20}) and (\ref{22}) together, 
 we obtain the expectation value as
\begin{align}
\langle Q\rangle = 
& \int \mathcal{D} \varphi_{+} \mathcal{D} \pi_{+} \mathcal{D} \varphi_{-} \mathcal{D} \pi_{-} 
 \varphi_{+}^{A_{1}}\left(\eta, \bm{x}_{1}\right) \cdots \varphi_{+}^{A_{N}}\left(\eta, \bm{x}_{N}\right) 
 \nonumber \\ 
& \times \exp \left[ i \int_{\eta_{0}}^{\eta_{f}} d \eta d^{3}x\,\left(\pi_{+A} \varphi_{+}^{\prime A}
 -\mathscr{H}\left[\pi_{+}, \varphi_{+}\right]\right)\right] 
 \nonumber \\ 
& \times \exp \left[- i \int_{\eta_{0}}^{\eta_{f}} d \eta d^{3} x\,\left(\pi_{-A} \varphi_{-}^{\prime A}
 -\mathscr{H}\left[\pi_{-}, \varphi_{-}\right]\right)\right] 
 \nonumber \\ 
& \times\left\langle\Omega | \varphi_{-}\left(\eta_{0}\right)\right\rangle 
 \left\langle\varphi_{+}\left(\eta_{0}\right) | \Omega\right\rangle 
 \prod_{A, \bm{x}} \delta\left(\varphi_{+}^{A}\left(\eta_{f}, \bm{x}\right)
 -\varphi_{-}^{A}\left(\eta_{f}, \bm{x}\right)\right),
\end{align}
where the path integral is restricted between the two times $\eta = \eta_{0}$ and $\eta = \eta_{f}$, 
 which means that it must be integrated over all possible states 
 $\ket{\varphi_{-}(\eta_{0})}$ and $\bra{\varphi_{+}(\eta_{0})}$ appearing in the integrand. 
As a result, two copies of the path integral were obtained. 
One is the  path integral with the direction in which time moves forward, 
 and the other is that with the the direction in which time moves backward, 
 both of which coincide in the limit of future time $\eta_{f}$ by condition
\begin{equation}
\varphi_{+}^{A}(\eta_{f}) = \varphi_{-}^{A}(\eta_{f}).
\end{equation}
If the theory does not contain higher-order derivatives, 
 then the momentum integrals can be performed as Gaussian integrals, 
 which can be written by using Lagrangian $\mathscr{L}_{\mathrm{cl}}\left[\varphi_{\pm}\right]$ as 
\begin{align}
\langle Q\rangle=
& \int \mathcal{D} \varphi_{+} \mathcal{D} \varphi_{-} \varphi_{+}^{A_{1}}\left(\eta, \bm{x}_{1}\right) 
 \cdots \varphi_{+}^{A_{N}}\left(\eta, \bm{x}_{N}\right) 
 \exp \left[ i \int_{\eta_{0}}^{\eta_{f}} d \eta d^{3} x\,\left(\mathscr{L}_{\mathrm{cl}}\left[\varphi_{+}\right]
 -\mathscr{L}_{\mathrm{cl}} \left[\varphi_{-}\right] \right)\right] 
 \nonumber  \\ 
& \times\left\langle\Omega | \varphi_{-}\left(\eta_{0}\right)\right\rangle
 \left\langle\varphi_{+}\left(\eta_{0}\right) | \Omega\right\rangle 
  \prod_{A, \bm{x}} \delta\left(\varphi_{+}^{A}\left(\eta_{f}, \bm{x}\right)-\varphi_{-}^{A}\left(\eta_{f}, \bm{x}\right)\right) .
\end{align}
Comparing to the path integral in flat space-time, 
 there is an extra contribution of factor $-i\mathscr{L}_{\mathrm{cl}}\left[\varphi_{-}\right]$. 
Since this is a change of $i$ to $-i$ in the path integral of the flat space-time, 
 it means that we only need to add the complex conjugate. 
The reason why such a complex conjugate of the original path integral is introduced is 
 that the expectation value of the operators in cosmology is determined at a fixed time, 
 which is different from that in particle physics 
 where the initial state is at infinite past and the final state is at infinite future. 
In addition, the two inner products $\left\langle\Omega | \varphi_{-}\left(\eta_{0}\right)\right\rangle$ 
 and $\left\langle\varphi_{+}\left(\eta_{0}\right) | \Omega\right\rangle$ play a role 
 in giving the correct $i\epsilon$ prescription of the time integrals. 
This means that the time-ordered part is used to deform the complex plane in the time direction 
\begin{equation}
\eta \to (1 - i \epsilon)\eta ,
\end{equation}
and the anti-time-ordered part is used to deform the complex plane in the time direction
\begin{equation}
\eta \to (1 + i \epsilon)\eta.
\end{equation}
Such a formalism, which is used to calculate the expectation values of physical quantities in non-equilibrium systems, 
 is called in-in formalism (Schwinger-Keldysh formalism). 
Corresponding to the extension of the path integral as described above, 
 the generating functional is extended as follows. 
For simplicity, we consider the case of a real scalar field $\varphi$, 
 but it is straightforward to generalize. 
We write $\varphi_{\pm}(\eta, \bm{x})$ for the two copies of the scalar field 
 and introduce their corresponding sources $J_{\pm}(\eta, \bm{x})$. 
\begin{equation}
Z\left[J_{+}, J_{-}\right]=\int \mathcal{D} \varphi_{+} \mathcal{D} \varphi_{-} 
 \exp \left[i \int_{\eta_{0}}^{\eta_{f}} \mathrm{d} \eta d^{3}x 
 \left(\mathscr{L}_{\mathrm{cl}}\left[\varphi_{+}\right]-\mathscr{L}_{\mathrm{cl}}\left[\varphi_{-}\right]
 +J_{+} \varphi_{+}-J_{-} \varphi_{-}\right)\right].
\end{equation}
We divide the Lagrangian into a free field part $\mathscr{L}_{0}$ 
 and an interaction part $\mathscr{L}_{\mathrm{int}}$,
\begin{equation}
\mathscr{L}_{\mathrm{cl}}[\varphi]=\mathscr{L}_{0}[\varphi]+\mathscr{L}_{\mathrm{int}}[\varphi]
\end{equation}
and rewrite the generating functional as 
\begin{eqnarray}
Z\left[J_{+}, J_{-}\right] 
&=& \exp \left[i \int_{\eta_{0}}^{\eta_{f}} d \eta d^{3}x
 \left(\mathscr{L}_{\mathrm{int}}\left[\frac{\delta}{i \delta J_{+}}\right]
 -\mathscr{L}_{\mathrm{int}}\left[-\frac{\delta}{i \delta J_{-}}\right]\right)\right] Z_{0}\left[J_{+}, J_{-}\right], \\ 
Z_{0}\left[J_{+}, J_{-}\right] & \equiv &\int \mathcal{D} \varphi_{+} \mathcal{D} \varphi_{-} 
 \exp \left[i \int_{\eta_{0}}^{\eta_{f}} d \eta d^{3}x\left(\mathscr{L}_{0}\left[\varphi_{+}\right]
 -\mathscr{L}_{0}\left[\varphi_{-}\right]+J_{+} \varphi_{+}-J_{-} \varphi_{-}\right)\right]. \nonumber \\
&{}& 
\end{eqnarray}
The path integral $Z_{0}[J_{+}, J_{-}]$ is Gaussian and can be explicitly integrated.
Therefore, there are four types of propagators, 
\begin{equation}
-i G_{a b}\left(\eta_{1}, \bm{x}_{1} ; \eta_{2}, \bm{x}_{2}\right) \equiv 
 \left.\frac{\delta}{i a \delta J_{a}\left(\eta_{1}, \bm{x}_{1}\right)} 
 \frac{\delta}{i b \delta J_{b}\left(\eta_{2}, \bm{x}_{2}\right)} Z_{0}\left[J_{+}, J_{-}\right]\right|_{J_{\pm}=0}.
\end{equation}
Since the signs of derivatives of $+$ type fields are different from those of $-$ type fields, 
 we have added $a, b=\pm$ to each derivative.
For example, the propagator of type $(+, +)$ is
\begin{eqnarray}
\label{eq287}
-i G_{++}\left(\eta_{1}, \bm{x}_{1} ; \eta_{2}, \bm{x}_{2}\right) 
&=& \left.\frac{\delta}{i \delta J_{+}\left(\eta_{1}, \bm{x}_{1}\right)} 
 \frac{\delta}{i \delta J_{+}\left(\eta_{2}, \bm{x}_{2}\right)} 
 Z_{0}\left[J_{+}, J_{-}\right]\right|_{J_{\pm}=0} \nonumber \\
&=& \int \mathcal{D} \varphi_{+} \mathcal{D} \varphi_{-} 
 \varphi_{+}\left(\eta_{1}, \bm{x}_{1}\right) \varphi_{+}\left(\eta_{2}, \bm{x}_{2}\right) 
 e^{i \int d \eta d^{3}x\left(\mathscr{L}_{0}\left[\varphi_{+}\right]-\mathscr{L}_{0}\left[\varphi_{-}\right]\right)} 
 \nonumber \\
&=& \left\langle\Omega\left|\mathrm{T}\left\{\varphi\left(\eta_{1}, \bm{x}_{1}\right) 
 \varphi\left(\eta_{2}, \bm{x}_{2}\right)\right\}\right| \Omega\right\rangle,
\end{eqnarray}
where T means an ordinary time-ordered operation.  
Thanks to the translational and rotational invariances on each time slice, 
 we can move to three-dimensional momentum space. 
The field $\varphi$ is represented by a mode function $v(\eta, \bm{k})$ 
 and creation and annihilation operators for a given three-dimensional momentum $\bm{k}$. 
Substituting this into the four propagators, we obtain the propagators in momentum space 
\begin{equation}
\Delta_{a b}\left(\eta_{1}, \eta_{2}, k\right)
=-i \int d^{3}x e^{-i\bm{k} \cdot \bm{x}} G_{a b}\left(\eta_{1}, \bm{x} ; \eta_{2}, \bm{0}\right) .
\end{equation}
Here, we added $-i$ to $\Delta_{ab}$ in order to avoid extra factors 
 in the Feymnan rule in momentum space. 
Furthermore, the momentum dependence of $\Delta_{ab}$ can be written as $k=|\bm{k}|$. 
This is because the propagator does not depend on the direction of the three-dimensional momentum, 
 but only on its magnitude, thanks to the rotational symmetry. 
Then, the tree-level propagators in three-dimensional momentum space can be easily obtained. 
\begin{eqnarray}
\Delta_{++}\left(\eta_{1}, \eta_{2}, k\right) 
&=& \Delta_{>}\left(\eta_{1}, \eta_{2}, k\right) \theta\left(\eta_{1}-\eta_{2}\right)
 +\Delta_{<}\left(\eta_{1}, \eta_{2}, k\right) \theta\left(\eta_{2}-\eta_{1}\right),  \\
\Delta_{+-}\left(\eta_{1}, \eta_{2}, k\right)&=&\Delta_{<}\left(\eta_{1}, \eta_{2}, k\right),  \\
\Delta_{-+}\left(\eta_{1}, \eta_{2}, k\right)&=& \Delta_{>}\left(\eta_{1}, \eta_{2}, k\right),  \\
\Delta_{--}\left(\eta_{1}, \eta_{2}, k\right)
&=& \Delta_{<}\left(\eta_{1}, \eta_{2}, k\right) \theta\left(\eta_{1}-\eta_{2}\right)
 +\Delta_{>}\left(\eta_{1}, \eta_{2}, k\right) \theta\left(\eta_{2}-\eta_{1}\right),
\end{eqnarray}
where $\Delta_{>}\left(\eta_{1}, \eta_{2}, k\right)$ and $\Delta_{<}\left(\eta_{1}, \eta_{2}, k\right)$ 
 are defined by
\begin{eqnarray}
\label{delta>}
&&\Delta_{>}\left(\eta_{1}, \eta_{2}, k\right)  \equiv v\left(\eta_{1}, k\right) v^{*}\left(\eta_{2}, k\right) , \\ 
&&\Delta_{<}\left(\eta_{1}, \eta_{2}, k\right) \equiv v^{*}\left(\eta_{1}, k\right) v\left(\eta_{2}, k\right)
\label{delta<}
\end{eqnarray}
and $\theta(\eta)$ is a step function of $\eta$. 
For the graphical representations of the propagators, 
 we assign black and white dots to represent $+$ and $-$, respectively. 
Hence, the four propagators are shown in Fig. \ref{Fig1}.
The external line connecting the time slice of the end time $\eta = \eta_{f}$ (boundary point) 
 adopts $\eta_{f}$ as an argument. 
Since the boundary point does not distinguish $+$ from $-$, 
 there are only two kinds of propagators, which are called as bulk-to-boundary propagators. 
We will assign a square to the boundary point as shown in Fig. \ref{Fig2}.
With these preparations, 
 let us derive the propagators of the $X, Y$ gauge bosons in the in-in formalism. 
The action of the $X, Y$ gauge fields (\ref{XYaction}) in the background space-time
\begin{equation}
ds^{2} = a^{2}(\eta)\left( -d\eta^{2} + dx^{2} + dy^{2} + dz^{2} \right), \quad a(\eta) = -\frac{1}{H\eta}
\end{equation}
can be written as
\begin{equation}
S_{G} =  \int d^{4}x 
\left[ 
\eta^{\mu\rho}X^{a*}_{\sigma}\partial_{\mu}\partial_{\rho}X^{\sigma}_{a} 
-X^{\mu*}_{a}\partial_{\mu}\partial_{\sigma}X^{\sigma}_{a}  
- a^{2}m_{X}^{2}X^{a}_{\mu}X^{\mu*}_{a} 
+ \frac{1}{\xi} X^{\rho*}_{a}\partial_{\mu}\partial_{\rho}X^{a\mu}  
\right] .
\end{equation}
Taking the Feynman gauge $\xi=1$, the second and fourth terms are cancelled each other 
 and we get a simple expression
\begin{equation}
S_{G}  =  \int d^{4}x 
\left[ \eta^{\mu\rho}X^{a*}_{\sigma}\partial_{\mu}\partial_{\rho}X^{\sigma}_{a} 
- a^{2}m_{X}^{2}X^{a}_{\mu}X^{\mu*}_{a} 
\right] .
\end{equation}
Deriving the equation of motion from this action, 
 we obtain the Klein-Gordon equation with mass $a^{2}m_{X}^{2}$ 
\begin{equation}
\left( \eta^{\nu\sigma}\partial_{\nu}\partial_{\sigma} - a^{2}m_{X}^{2} \right) X^{\mu}_{a} =0.
\label{KKeq}
\end{equation}
If we Fourier transform the gauge field $X^{\mu}_{a}$ into three dimensional momentum space
 and write its mode function as $v^{\mu}_{a}(\eta, \bm{k})$, 
 the Klein-Gordon equation (\ref{KKeq}) is
\begin{equation}
\frac{\partial^{2}v^{\mu}_{a}}{\partial \eta^{2}}(\eta, \bm{k}) 
 + \left(|\bm{k}|^{2} + \frac{m_{X}^{2}}{H^{2}\eta^{2}} \right)v^{\mu}_{a}(\eta, \bm{k}) = 0 .
\end{equation}
If we define 
\begin{equation}
\nu_{A} \equiv \sqrt{\frac{1}{4} - \left(   \frac{m_{X}}{H} \right)^{2}  } ,
\end{equation}
the solution to this Klein-Gordon equation is found to be 
\begin{equation}
v^{(\lambda)\mu}_{a}(\eta, \bm{k}) 
= -i\frac{\sqrt{\pi}}{2}e^{i\pi\left( \frac{\nu_{A}}{2} +\frac{1}{4}   \right)} 
 \left( -\eta  \right)^{\frac{1}{2}} H^{a(1)}_{\nu_{A}}(-|\bm{k}|\eta)
 \varepsilon^{(\lambda)\mu}(\bm{k})\qquad (\lambda = 1, 2, 3) ,
\label{Hankel}
\end{equation}
where $ H^{a(1)}_{\nu_{A}}(-|\bm{k}|\eta)$ is a Hankel function of the first kind 
 and the index $a$ is added as the gauge field index. 
$\varepsilon^{(\lambda)\mu}(k)$ is a polarization vector of $X_\mu$, 
 which satisfies the normalization condition and the projection relation
\begin{equation}
\varepsilon^{(\lambda)\mu}(\bm{k})\varepsilon^{(\lambda)\nu}(\bm{k}) 
 = \delta^{\lambda \lambda'}\delta^{\mu\nu}, 
 \qquad \sum_{\lambda} \varepsilon^{(\lambda)}_{\mu}(\bm{k})\varepsilon_{\nu}^{(\lambda)}(\bm{k}) 
 = g_{\mu\nu} + \frac{k_{\mu}k_{\nu}}{m_{X}^{2}},
\end{equation}
where $\lambda$ expresses a polarization index. 
The normalization of the expression (\ref{Hankel}) was done as follows. 
First, the Bunch-Davies vacuum is chosen as the initial condition 
 and the gauge field $X^{\mu}_{a}(\eta, \bm{x})$ and its conjugate momentum $\Pi^{\mu}_{a}(\eta, \bm{x})$ 
 is expanded in terms of the creation and annihilation operators. 
\begin{eqnarray}
X^{\mu}_{a}(\eta, \bm{x}) &=& \int\frac{d^{3}k}{(2\pi)^{3}} 
 \sum_{\lambda=1, 2, 3}\left[ a^{(\lambda)}(\bm{k})v^{(\lambda)\mu}_{a}(\eta, \bm{k}) 
  + b^{(\lambda)\dagger}(-\bm{k})v^{(\lambda)\mu*}_{a}(\eta, -\bm{k}) \right] e^{i\bm{k}\cdot \bm{x}}, \\
\Pi^{\mu}_{a}(\eta, \bm{x}) &=& \frac{\partial}{\partial \left(\partial_{\eta}{X_{a\mu}}\right)} 
 \left[ \eta^{\mu\rho}X^{a*}_{\sigma}\partial_{\mu}\partial_{\rho}X^{\sigma}_{a} 
 - a^{2}m_{X}^{2}X^{a}_{\mu}X^{\mu*}_{a} \right]
 \nonumber \\
&=& \int\frac{d^{3}k}{(2\pi)^{3}} \sum_{\lambda=1, 2, 3} 
 \left[ b^{(\lambda)}(-\bm{k})\partial_{\eta}v^{(\lambda)\mu}_{a}(\eta, -\bm{k}) 
+ a^{(\lambda)\dagger}(\bm{k}) 
\partial_{\eta} v^{(\lambda)\mu*}_{a}(\eta, \bm{k})\right] e^{-i\bm{k}\cdot \bm{x}}, 
\nonumber \\
\end{eqnarray}
where $a^{(\lambda)}(\bm{k}), a^{(\lambda)\dagger}(\bm{k}), b^{(\lambda)}(-\bm{k})$ 
 and $b^{(\lambda)\dagger}(-\bm{k})$ are the creation and  annihilation operators 
 corresponding to the modes with a polarization $\lambda$. 
Then, the gauge field, its conjugate momentum and the creation, annihilation operators 
 are set to satisfy the following canonical commutation relations
\begin{eqnarray}
\left[X^{\mu}_{a}(\eta, \bm{x}), \Pi^{\nu}_{a}(\eta, \bm{x}') \right] 
 &=& i \left( g^{\mu\nu} - \frac{\partial^{\mu} \partial^{\nu}}{m_{X}^{2}} \right) \delta( \bm{x} - \bm{x}' ), \\
\left[a^{(\lambda)}(\bm{k}), a^{(\lambda')\dagger}(\bm{k}') \right] 
 &=& (2\pi)^{3}\delta^{\lambda \lambda'} \delta(\bm{k} - \bm{k}'), \\
\left[b^{(\lambda)}(\bm{k}), b^{(\lambda')\dagger}(\bm{k}') \right] 
  &=& (2\pi)^{3}\delta^{\lambda \lambda'} \delta(\bm{k} - \bm{k}'), \\
\text{others} &=& 0. 
\end{eqnarray}
Using (\ref{delta>}), (\ref{delta<}) and a property
\begin{equation}
H_{\nu_{A}}^{(1)*}(-k\eta) = H_{\nu_{A}}^{(2)}(-k\eta)
\end{equation}
that the complex conjugate of Hankel functions of the first kind is a Hankel function of the second kind, 
 we obtain the propagators of the $X, Y$ gauge bosons, 
\begin{eqnarray}
&&\Delta^{(\lambda)\mu\nu}_{>}\left(\eta_{1}, \eta_{2}, k\right)  
= -\frac{\pi}{4} e^{-\pi \mathrm{Im}(\nu_{A})} \left(  \eta_{1} \eta_{2}\right)^{1/2} 
H_{\nu_{A}}^{(1)}(-k\eta_{1}) H_{\nu_{A}}^{(2)}(-k\eta_{2}) \varepsilon^{(\lambda)\mu}(k) \varepsilon^{(\lambda)\nu}(k), \\
&&\Delta^{(\lambda)\mu\nu}_{<}\left(\eta_{1}, \eta_{2}, k\right) 
= -\frac{\pi}{4} e^{-\pi \mathrm{Im}(\nu_{A})} \left(  \eta_{1} \eta_{2}\right)^{1/2} 
H_{\nu_{A}}^{(1)}(-k\eta_{2}) H_{\nu_{A}}^{(2)}(-k\eta_{1}) \varepsilon^{(\lambda)\nu}(k) \varepsilon^{(\lambda)\mu}(k) .
\end{eqnarray}

\subsection{Interaction of $X, Y$ gauge bosons and gravitons}
For the calculation of the non-Gaussianity from $X, Y$ gauge bosons, 
 let us calculate the interactions between $X, Y$ gauge bosons and gravitons. 
Although we have to calculate the interactions between the $X, Y$ gauge bosons and the inflaton 
 to predict the non-Gaussianity, they are very complicated. 
Therefore, we will instead calculate three-point function of the gravition 
 and evaluate the three-point function of the inflation 
 from the results of three-point function of the gravition 
 by utilizing information on the power spectrum. 
As we will see later, this strategy is sufficient for our analysis to 
estimate the order of non-Gaussianity. 
We proceed with the calculation using the action of the gauge field (\ref{XYaction}) 
 and the inverse matrix of the metric (\ref{inversemetric}). 
The following three-point and four-point couplings including the gravition are derived, respectively.
\begin{eqnarray}
\mathcal{L}_{3pt} &=& \gamma^{jl}  
\left[  
 -   a \left( \partial_{j} X_{0}  \right) \partial_{l} X^{*}_{0} 
+ a \left( \partial_{j} X_{0}  \right) \partial_{0} X^{*}_{l} + a \left( \partial_{0} X_{j}  \right) \partial_{l} X^{*}_{0} 
- a \left( \partial_{j} X_{l}  \right) \partial_{0} X^{*}_{0} \right. \nonumber \\
&& \left. - a \left( \partial_{0} X_{0}  \right) \partial_{l} X^{*}_{l}   
- a \left( \partial_{0} X_{j}  \right) \partial_{0} X^{*}_{l}  
 +   \frac{1}{a}\left( \partial_{n} X_{j}  \right) \partial_{n} X^{*}_{l} 
 + \frac{1}{a}\left( \partial_{j} X_{n}  \right) \partial_{l} X^{*}_{n} \right. \nonumber \\
&& \left. - \frac{1}{a}\left( \partial_{n} X_{j}  \right) \partial_{l} X^{*}_{n} 
 -  \frac{1}{a}\left( \partial_{j} X_{n}  \right) \partial_{n} X^{*}_{l} +  \frac{1}{a}\left( \partial_{n} X_{n}  \right) \partial_{j} X^{*}_{l} 
+  \frac{1}{a}\left( \partial_{j} X_{l}  \right) \partial_{n} X^{*}_{n}  \right. \nonumber \\
&& - \left.  am_{X}^{2} X_{j} X_{l}^{*} \right], 
\label{3pt} \\
%
\mathcal{L}_{4pt} &=& \frac{1}{2} \gamma^{im} \gamma^{mk}  
\left[ 
a \left( \partial_{i} X_{0}  \right) \partial_{k} X^{*}_{0} 
- a \left( \partial_{i} X_{0}  \right) \partial_{0} X^{*}_{k} - a \left( \partial_{0} X_{i}  \right) \partial_{k} X^{*}_{0} 
 + a \left( \partial_{i} X_{k}  \right) \partial_{0} X^{*}_{0} 
 \right. \nonumber \\
&& \left.+ a \left( \partial_{0} X_{0}  \right) \partial_{i} X^{*}_{k}  
 + a \left( \partial_{0} X_{i}  \right) \partial_{0} X^{*}_{k}   
 -  \frac{1}{a}\left( \partial_{n} X_{i}  \right) \partial_{n} X^{*}_{k} 
-  \frac{1}{a}\left( \partial_{i} X_{n}  \right) \partial_{k} X^{*}_{n} 
\right.  \nonumber \\
&& \left. +  \frac{1}{a}\left( \partial_{n} X_{i}  \right) \partial_{k} X^{*}_{n} 
 +  \frac{1}{a}\left( \partial_{i} X_{n}  \right) \partial_{n} X^{*}_{k}  
 -  \frac{1}{a}\left( \partial_{n} X_{n}  \right) \partial_{i} X^{*}_{k} 
-  \frac{1}{a}\left( \partial_{i} X_{k}  \right) \partial_{n} X^{*}_{n} 
\right. \nonumber \\
&& + \left.  a m_{X}^{2} X_{i}  X^{*}_{k}  \right] 
+ \frac{1}{a} \gamma^{im} \gamma^{kn}   
\left[ -  \left( \partial_{i} X_{k}  \right) 
\partial_{m} X^{*}_{n} + \left( \partial_{i} X_{k}  \right) \partial_{n} X^{*}_{m} 
- \left( \partial_{i} X_{m}  \right) \partial_{k} X^{*}_{n} \right]. 
\label{4pt}
\end{eqnarray}
For the sake of clarity, we omitted the subscript $a$ for the gauge field.

\section{Calculation of non-Gaussianity}
\subsection{Contribution from $X, Y$ gauge bosons to the graviton 3-point function}
\begin{figure}[htbp]
 \begin{center}
  \includegraphics[width=100mm]{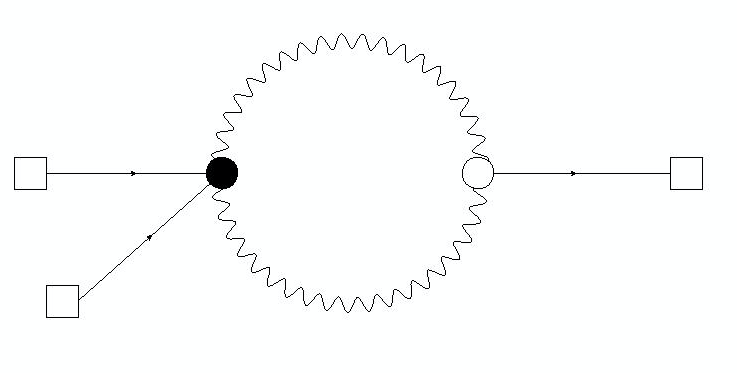}
 \end{center}
 \vspace*{-5mm}
 \caption{($+, -$) type graph of the contribution from $X, Y$ gauge bosons 
 to the graviton three-point function.}
 \label{Fig3}
 \end{figure}

We are ready to calculate one-loop contributions from $X, Y$ gauge bosons 
 to the non-Gaussianity for the graviton three-point function. 
Note that the terms multiplied by $a^{-1}$ among the terms 
 in the three-point and four-point couplings (\ref{3pt}), (\ref{4pt})
 become small after the inflation, because of the scale factor $a \propto e^{Ht} \to \infty$ as $t \to \infty$. 
Therefore, it is good approximation to neglect the terms proportional to $1/a$. 
The remaining interactions we should consider are classified into two types. 
One is the interactions from the derivative terms for the $X, Y$ gauge bosons, 
 the other is those from the mass terms for them. 
We first consider the interaction like
\begin{equation}
-\gamma^{jl} a \left( \partial_{j} X_{0}  \right) \partial_{l} X^{*}_{0} 
\times \frac{1}{2} \gamma^{im} \gamma^{mk}  a \left( \partial_{i} X_{0}  \right) \partial_{k} X^{*}_{0}, 
\end{equation}
which are the first term of the three- and four-point couplings (\ref{3pt}) and (\ref{4pt}), respectively, 
 and calculate the non-Gaussianity of the graviton three point functions as a typical example by using these interactions. 
The Feynman diagram for this is shown in Figure \ref{Fig3}. 
The external gravitons are represented by solid lines, 
 and the internal $X, Y$ gauge bosons are represented by wavy lines. 

Using the Feynman rules in the in-in formalism, 
 the contribution of this one-loop diagram to the graviton three-point function can be written as
\begin{align}
& S \times N \times \int_{-(1+i\epsilon)\infty}^{0}a(\tau) \, d\tau \int_{-(1+i\epsilon)\infty}^{0}a(\tau') \, d\tau' 
 \int \frac{d^{3}k} {(2\pi)^{3}} i \cdot (-i) \frac{1}{2} a(\tau)  \left(  -a(\tau')  \right) 
  \nonumber \\
& \times  G_{++}^{ab, im}(\tau , 0, \bm{p}_{1}) G_{++}^{bc, mk}(\tau , 0, \bm{p}_{2}) G_{++}^{ac, jl}(\tau' , 0, \bm{p}_{3}) 
\nonumber \\
& \times \frac{1}{3} \sum_{\lambda} \Delta_{00 +-}^{(\lambda)} ( \tau, \tau', k) \cdot \left( -ik_{i}  \right) \left( -ik_{j}  \right)  
\nonumber \\
& \times\frac{1}{3} \sum_{\lambda'} \Delta_{00 +-}^{(\lambda')} 
\left( \tau, \tau', \left| \bm{k} - \bm{p}_{1} - \bm{p}_{2} \right| \right) 
\cdot \left( -i \left(  \bm{k} - \bm{p}_{1} - \bm{p}_{2} \right)_{k}  \right) 
\left( -i  \left(  \bm{k} - \bm{p}_{1} - \bm{p}_{2} \right)_{l} \right) 
\nonumber \\
& + \text{sym.} + \text{c.c.} \\
 \equiv & J +  \text{sym.} + \text{c.c.},
\label{1-loop1}
\end{align}
where c.c. means a complex conjugate of the previous terms. 
Here, we write the first term as $J$ and 
 ``sym.'' represents contributions from the diagrams 
 with cyclically permutated momenta of $\bm{p}_{1}, \bm{p}_{2}$ and $\bm{p}_{3}$.  
The momentum flowing into the black circles is denoted by $\bm{p}_{1}$ and $\bm{p}_{2}$, 
 the momentum leaving the white circles denoted by $\bm{p}_{3}$, 
 and the internal line momenta are denoted by $\bm{k}$ and $\bm{k}-\bm{p}_{1}-\bm{p}_{2}$. 
$S$ is a symmetry factor, which is the number of ways 
 to connect legs of the vertex to the external lines, 
 namely, $2^2 \times 2$ in the present case.  
$N$ is a dimension of $X, Y$  gauge bosons. 
In an $SU(5)$ GUT case, $N=3 \times 2$ 
 because $X, Y$ gauge bosons belong to the $({\bf 3}, {\bf 2})$ and $({\bf 3}^*, {\bf 2})$ representations 
 under $SU(3) \times SU(2)$ of the SM gauge groups. 
$G_{++}^{ab, im}(\tau , 0, \bm{p}_{1})$ and so on are graviton propagators \cite{SK}
\begin{equation}
\left(G_{>} \right)_{ij, kl}(\tau_{1}, \tau_{2}, \bm{k}) 
= \frac{H^{2}} {M_{\mathrm{Pl}}^{2}k^{3}} \left( 1+ ik\tau_{1}   \right) 
\left( 1- ik\tau_{2}   \right) e^{-ik \left(   \tau_{1} - \tau_{2} \right)} 
\sum_{\alpha} e^{\alpha}_{ij} (\bm{k}) e^{\alpha *}_{kl} (\bm{k}).
\end{equation}
Here only the results of Eq.~(\ref{1-loop1}) are given because of complicated calculations. 
For detailed calculations, see Appendix A. 
\begin{equation}
\begin{aligned}
\text{Eq.}~(\ref{1-loop1}) \simeq &~SN \frac{1}{18\cdot 4^{3} \pi} \frac{H^{2}} { M_{\mathrm{Pl}}^{6} (p_{1}p_{2}p_{3}  )^{3} } 
\Pi_{abim}(\bm{p}_{1}) \Pi_{bcmk}(\bm{p}_{2}) \Pi_{acjl}(\bm{p}_{3}) \\
& \times \frac{1}{210\pi} \left(  \delta_{ij}  \delta_{kl} +  \delta_{ik}  \delta_{jl} +  \delta_{il}  \delta_{jk}   \right) 
\left(   2\Lambda - p_{1} - p_{2}  \right)^{7} + \text{sym.}
\end{aligned}
\label{1-loop1result}
\end{equation}
where $\Pi_{ijkl} (\bm{k})$ is defined as an expression for the sum of the projectors by the graviton polarization tensor
\begin{equation}
\label{eq3.4}
\Pi_{ ijkl} (\bm{k}) \equiv \sum_{\alpha} e^{\alpha}_{ij} (\bm{k}) e^{\alpha *}_{kl} (\bm{k}).
\end{equation}
In the result (\ref{1-loop1result}), the large momentum limit for Hankel function 
 and the squeezed limit for external momenta are considered. 
$\Lambda$ is the cutoff scale of the loop momentum, 
 which is usually considered to be $m_X$ from the effective theory viewpoint.  
If the cutoff scale is sufficiently larger than the graviton momentum, then Eq.~(\ref{1-loop1result}) can be
\begin{equation}
\begin{aligned}
\text{Eq.}~(\ref{1-loop1result}) \simeq &~SN \frac{1}{18\cdot 4^{3} \pi} 
\frac{H^{2}} { M_{\mathrm{Pl}}^{6} (p_{1}p_{2}p_{3}  )^{3} } \Pi_{abim}(\bm{p}_{1}) 
\Pi_{bcmk}(\bm{p}_{2}) \Pi_{acjl}(\bm{p}_{3}) \\
& \times \frac{1}{210\pi} \left(  \delta_{ij}  \delta_{kl} +  \delta_{ik}  \delta_{jl} +  \delta_{il}  \delta_{jk}   \right) 
\left(   2\Lambda  \right)^{7} \times 3.
\end{aligned}
\label{1-loop1results2}
\end{equation}
Similarly, we have to calculate graphs with propagators 
 other than those of type $(+, -)$  
 and found that they are of the same order as the result of (\ref{1-loop1results2}). 
Also, the one-loop contributions with other terms proportional to $a$ 
 in the three-point couplings (\ref{3pt}) and four-point couplings (\ref{4pt}) to the graviton three-point function 
 are easily found to be the same order as the results (\ref{1-loop1results2}).
The remaining last contribution due to the three-point and four-point couplings 
  originated from the mass terms for $X, Y$ gauge bosons is expressed as
\begin{equation}
 - \gamma^{jl}   am_{X}^{2} X_{j} X_{l}^{*} \times  \frac{1}{2} \gamma^{im}  a m_{X}^{2} X_{i}  X^{*}_{k}\gamma^{mk}. 
\end{equation}
The corresponding one-loop diagram such as Fig.~\ref{Fig3} is similarly calculated (denote by $J_{m}$),
\begin{equation}
J_{m} \simeq \text{eq.}(\ref{1-loop1}) \times 2^{4} \times 30 \times \left(  \frac{m_{X}}{H} \right)^{4} .
\label{1-loop2}
\end{equation}
As is clear from the result, 
 the contribution $J_m$ is dominated compared to $J$ 
 due to the enhancement factors $2^{4} \times 30 \times \left(  \frac{m_{X}}{H} \right)^{4}$. 
In particular, the factor $\left(  \frac{m_{X}}{H} \right)^{4}$ is very large 
 in the case where the mass of $X, Y$ gauge bosons is larger than the Hubble scale 
 and crucial in estimating the non-Gaussianity.

\subsection{Non-Gaussianity}

\begin{figure}[htbp]
 \begin{center}
  \includegraphics[width=16cm, bb=9 9 1248 181]{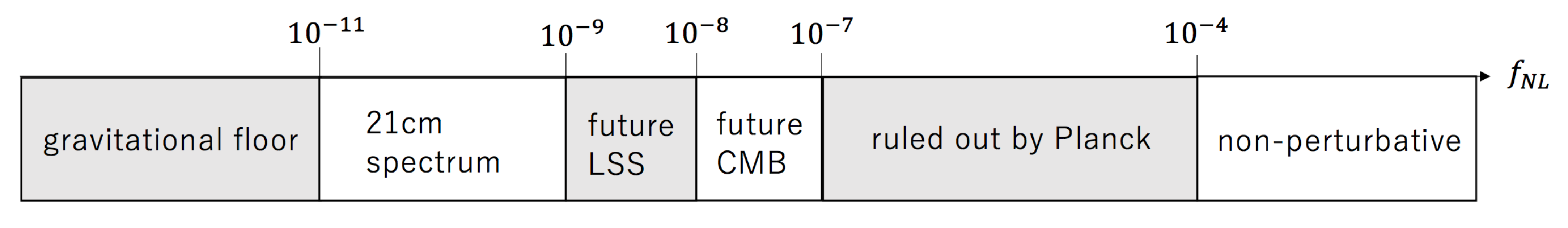}
 \end{center}
 \caption{Schematic illustration of current 
 and future constraints on the non-Gaussianity 
 (Based on a figure in \cite{Baumann2018}).}
 \label{Fig4}
 \end{figure}

To evaluate the Non-Gaussianity, we consider the graviton version of the equation (5.9) in \cite{Chen}. 
\begin{equation}
\left\langle \gamma \gamma \gamma \right\rangle \equiv S\left(p_{1}, p_{2}, p_{3}\right) 
 \frac{1}{\left(p_{1} p_{2} p_{3}\right)^{2}} \tilde{P}_{\gamma}^{2}(2 \pi)^{7} \delta^{3}
 \left(\sum_{i=1}^{3} \mathbf{p}_{i}\right), 
\end{equation}
where $\tilde{P}_{\gamma}$ is the reference power spectrum. 
Since $r \sim 0.1$ is the case where the tensor-to-scalar ratio is largest from the observed data, 
 using the power spectrum $\tilde{P}_{\zeta}= \tilde{P}_{\zeta} (k_{\mathrm{WMAP}}) =6.1 \times 10^{-9}$ 
 of the inflaton in $k_{\mathrm{WMAP}}=0.027\,\mathrm{Mpc}^{-1}$, we obtain
\begin{equation}
\tilde{P}_{\gamma}^{2} = r^{2} \tilde{P}_{\zeta}^2 \sim 3.7 \times 10^{-19}.
\label{Pgamma}
\end{equation}
Consider an equilateral triangle $p_{1}=p_{2}=p_{3}=p$ as the momentum configuration. 
Since the natural scale in the theory is the GUT scale $m_{X}$, 
 we write this momentum as $p=\alpha m_{X}$ with the arbitrary dimensionless parameter $\alpha$. 
Using the loop momentum cutoff as $\Lambda \simeq m_{X} = 10^{15}$ GeV, 
 Hubble scale $H=10^{14}$ GeV, Planck scale $M_{\mathrm{Pl}}=10^{19}$ GeV and equation (\ref{Pgamma}), 
 the function $S\left(p_{1}, p_{2}, p_{3}\right)$ can be evaluated as
\begin{eqnarray}
S \left(p_{1}, p_{2}, p_{3}\right) 
&=& \frac{ \left(p_{1} p_{2} p_{3}\right)^{2} }  { (2 \pi)^{7} \tilde{P}_{\gamma}^{2} } 
\times  \frac{SN}{36\cdot 4^{3} \pi} \frac{H^{2}} { M_{\mathrm{Pl}}^{6} (p_{1}p_{2}p_{3}  )^{3} } 
\Pi_{abim}(\bm{p}_{1}) \Pi_{bcmk}(\bm{p}_{2}) \Pi_{acjl}(\bm{p}_{3}) \nonumber \\
&{} &\times \frac{1}{210\pi} \left(  \delta_{ij}  \delta_{kl} +  \delta_{ik}  \delta_{jl} +  \delta_{il}  \delta_{jk}   \right) 
\left(   2\Lambda  \right)^{7} \times 3  \times 2^{4} \times 30 \times \left(  \frac{m_{X}}{H} \right)^{4} 
\nonumber \\
& \sim & \mathcal{O} (10^{-9}) \times \frac{1} { \alpha^{3} }.
\end{eqnarray}
In the limit $p_{1}=p_{2}=p_{3}$, 
 the non-Gaussianity $f_{\mathrm{NL}}$ is of the same order 
 as the function $S\left(p_{1}, p_{2}, p_{3}\right)$ \cite{Chen} 
 and the non-Gaussianity $f_{\mathrm{NL}}$ for the inflaton three-point function 
  is obtained by multiplying $S\left(p_{1}, p_{2}, p_{3}\right)$ 
  for the graviton three-point finction by $r^{3/2}$. 
\begin{equation}
f_{\mathrm{NL}} \sim  \mathcal{O} (10^{-11}) \times \frac{1} {\alpha^{3} } .
\label{NLGUT}
\end{equation}
Considering a larger GUT group such as an $E_8$ group, 
 we can expect the number of $X, Y$ gauge bosons contributing to the diagram to increase, 
 six times enhancement compared to $SU(5)$ in the case of $E_8$ for instance. 
In addition, we used in the above calculation the non-SUSY GUT scale $\simeq 10^{15}$ GeV, 
 but if we consider the SUSY GUT, then $m_{X}$ becomes typically one order of magnitude larger $10^{16}$ GeV 
 and the factor $\left(  m_{X}/H \right)^{4}$ makes $f_{\mathrm{NL}}$ 
 four orders of magnitude larger.\footnote{It is assumed that 
 the order estimation of our results are not so
 changed by the contributions from the superpartners. }  
In such a case, the non-Gaussianity is
\begin{equation}
f_{\mathrm{NL}} \sim  \mathcal{O} (10^{-7}) / \alpha^{3}.
\label{NLSUSYGUT}
\end{equation}
Alternatively, if we consider the string-inspired GUT, 
 the $m_X$ would be typically a string scale $10^{17}$ GeV, 
 then the factor $\left( m_{X}/H \right)^{4}$ makes $f_{\mathrm{NL}}$ eight orders of magnitude 
 larger\footnote{It is assumed that the order estimation of our results are not so
 changed by the contributions from the string modes and so on.}, 
 which results in
\begin{equation}
f_{\mathrm{NL}} \sim  \mathcal{O} (10^{-3}) / \alpha^{3}.
\label{NLstringGUT}
\end{equation} 
In order to discuss a detectability of the non-Gaussianity form $X, Y$ gauge bosons, 
 the schematic illustration of current and future constraints on the non-Gaussianity 
 is given in Figure \ref{Fig4}. 
The results depend on the parameter $\alpha$, 
 which normalizes the external momentum of the gravitions by $m_X$.  
From the effective theory viewpoint, 
 the external momentum should be smaller than $m_X$, 
 namely $\alpha \leq {\mathcal O}(1)$. 
For the non-SUSY GUT case (\ref{NLGUT}), 
 if the parameter $\alpha$ is in a range of ${\mathcal O(10^{-(1\sim 2)})} < \alpha < {\mathcal O}(1)$, 
 the non-Gaussianity from $X, Y$ gauge bosons may be verified 
 by future experiments of the CMB, observations of large-scale structures in the universe, 
 and observations of the 21cm spectrum of neutral hydrogen atoms. 
For the SUSY GUT case (\ref{NLSUSYGUT}), 
 $\alpha \sim {\mathcal O}(1)$ implies that the non-Gaussianity from $X, Y$ gauge bosons may be verified 
 by future experiments of the CMB. 
As for the string-inspire GUT case (\ref{NLstringGUT}), 
 our prediction is that it has already been ruled out by the Planck satellite experiments.  

\section{Conclusion}
Although the GUT is one of the attractive extensions of the SM, 
 it is very hard to detect its signatures directly at colliders 
 since the GUT scale is predicted to be very high. 
Typical predictions of GUT are a proton decay, 
 but it has not been observed at Super-Kamiokande so far. 
Recent developments of the precision measurements of CMB by the Planck satellites etc 
 has opened an alternative avenue to a possible detection of the GUT signatures. 
If the heavy fields of the GUT scale are present during the inflation, 
 which is considered to happen typically at an order of $10^{14}$ GeV close to the GUT scale, 
 the heavy fields will contribute to the three-point function of the inflaton. 
In order to evaluate such contributions, 
 we consider an approach based on the effective field theory of inflation. 

In this paper, we discussed a detactability of the $X, Y$ gauge boson in GUT, 
 which is model-independently present in GUT.  
We have first calculated one-loop contributions of $X, Y$ gauge bosons 
 to the graviton three-point function, which is based on the in-in formalism. 
Then, the obtained results were converted to the inflaton three-point functions 
 by using the information on the power spectrum. 
The remarkable feature found in this paper is that 
 one-loop diagram with the cubic and the quartic interactions 
 between the $X, Y$ gauge bosons and the gravitons, 
 which are originated from the mass terms of $X, Y$ gauge bosons, 
 provides an enhancement factor such as $(m_X/H)^4$. 
Due to this enhancement factor, 
 it might be possible to detect the non-Gaussianity from $X, Y$ gauge bosons 
 depending on the magnitude of external momentum of the graviton. 
For the non-SUSY GUT case, 
 if the parameter $\alpha$ is ${\mathcal O(10^{-(1 \sim 2)})} < \alpha < {\mathcal O}(1)$, 
 the non-Gaussianity from $X, Y$ gauge bosons may be verified 
 by future experiments of the CMB, observations of large-scale structures in the universe, 
 and observations of the 21cm spectrum of neutral hydrogen atoms. 
For the SUSY GUT case, 
 ${\mathcal O}(1)$ implies that the non-Gaussianity from $X, Y$ gauge bosons may be verified 
 by future experiments of the CMB. 
As for the string-inspire GUT case, 
 our prediction is that it has already been ruled out by the Planck satellite experiments.


\section*{Acknowledgments}
This work is supported in part by JSPS KAKENHI Grant Number JP17K05420 (N.M.).

\appendix
\section{Details of calculation}
In this appendix, we give a detailed calculation of the one-loop diagram for the graviton three-point function. 
The diagrams we should calculate are classified into two categories. 
One is the diagrams with a three-point and a four-point interactions from the derivative terms for the $X, Y$ gauge bosons, 
 the other is those from the mass terms for them. 

We first discuss former one, which is calculated from the sum of $J$ denoted in the main text and its complex conjugate.   
\begin{align}
\label{app1}
J =& SN   \int_{-(1+i\epsilon)\infty}^{0}a(\tau) \, d\tau \int_{-(1+i\epsilon)\infty}^{0}a(\tau') 
\, d\tau' \int \frac{d^{3}k} {(2\pi)^{3}} i \cdot (-i) \frac{1}{2} a(\tau)  \left(  -a(\tau')  \right) 
\nonumber \\
& \times  G_{++}^{ab, im}(\tau , 0, \bm{p}_{1}) 
G_{++}^{bc, mk}(\tau , 0, \bm{p}_{2}) G_{++}^{ac, jl}(\tau' , 0, \bm{p}_{3}) \nonumber \\
& \times \frac{1}{3} \sum_{\lambda} \Delta_{00 +-}^{(\lambda)} 
( \tau, \tau', k) \cdot \left( -ik_{i}  \right) \left( -ik_{j}  \right)  
\nonumber \\
& \times\frac{1}{3} \sum_{\lambda'} \Delta_{00 +-}^{(\lambda')} 
\left( \tau, \tau', \left| \bm{k} - \bm{p}_{1} - \bm{p}_{2} \right| \right) 
\cdot \left( -i \left(  \bm{k} - \bm{p}_{1} - \bm{p}_{2} \right)_{k}  \right) 
\left( -i  \left(  \bm{k} - \bm{p}_{1} - \bm{p}_{2} \right)_{l} \right). 
\end{align}
Using the graviton propagator \cite{SK}
\begin{equation}
\label{app2}
\left(G_{>} \right)_{ij, kl}(\tau_{1}, \tau_{2}, \bm{k}) = \frac{H^{2}} {M_{\mathrm{Pl}}^{2}k^{3}} 
\left( 1+ ik\tau_{1}   \right)  \left( 1- ik\tau_{2}   \right) e^{-ik \left(   \tau_{1} - \tau_{2} \right)} 
\sum_{\alpha} e^{\alpha}_{ij} (\bm{k}) e^{\alpha *}_{kl} (\bm{k}) 
\end{equation}
and the $X, Y$ gauge boson propagators
\begin{eqnarray}
\label{app3}
&&\Delta^{(\lambda)\mu\nu}_{>}\left(\eta_{1}, \eta_{2}, k\right) 
= -\frac{\pi}{4} e^{-\pi \mathrm{Im}(\nu_{A})} \left(  \eta_{1} 
\eta_{2}\right)^{1/2} 
H_{\nu_{A}}^{(1)}(-k\eta_{1}) H_{\nu_{A}}^{(2)}(-k\eta_{2}) 
\varepsilon^{(\lambda)\mu}(k) \varepsilon^{(\lambda)\nu}(k), \\
\label{app4}
 &&\Delta^{(\lambda)\mu\nu}_{<}\left(\eta_{1}, \eta_{2}, k\right) 
= -\frac{\pi}{4} e^{-\pi \mathrm{Im}(\nu_{A})} \left(  \eta_{1} \eta_{2}\right)^{1/2} 
H_{\nu_{A}}^{(1)}(-k\eta_{2}) H_{\nu_{A}}^{(2)}(-k\eta_{1}) 
\varepsilon^{(\lambda)\nu}(k) \varepsilon^{(\lambda)\mu}(k),
\end{eqnarray}
we obtain
\begin{align}
\label{app5}
J + \text{c.c.} = & 2 \mathrm{Re} \left[ \frac{SN}{18} 
\int_{-(1+i\epsilon)\infty}^{0} \, d\tau \int_{-(1+i\epsilon)\infty}^{0} \, d\tau' 
\int \frac{d^{3}k} {(2\pi)^{3}} \left( -\frac{1}{H\tau}  \right)^{2} \left( -\frac{1}{H\tau'}  \right)^{2} \right.
\nonumber \\
& \times \left[ \frac{H^{2}} {M_{\mathrm{Pl}}^{2}p_{1}^{3}}  
\left( 1+ ip_{1}\tau  \right) e^{-ip_{1}   \tau} 
\sum_{\alpha} e^{\alpha}_{ab} (\bm{p}_{1}) e^{\alpha *}_{im} (\bm{p}_{1}) \theta(\tau) \right. 
\nonumber \\
 & \left. \quad + \frac{H^{2}} {M_{\mathrm{Pl}}^{2}p_{1}^{3}} 
 \left( 1- ip_{1}\tau  \right) e^{ip_{1}   \tau} \sum_{\alpha} e^{\alpha *}_{ab} (\bm{p}_{1}) 
 e^{\alpha }_{im} (\bm{p}_{1}) \theta(-\tau)\right]  
 \nonumber \\
& \times \left[ \frac{H^{2}} {M_{\mathrm{Pl}}^{2}p_{2}^{3}} 
\left( 1+ ip_{2}\tau  \right) e^{-ip_{2} \tau} \sum_{\beta} 
e^{\beta}_{bc} (\bm{p}_{2}) e^{\alpha *}_{mk} (\bm{p}_{2}) \theta(\tau) \right. 
\nonumber \\
& \left. \quad + \frac{H^{2}} {M_{\mathrm{Pl}}^{2}p_{2}^{3}} 
\left( 1- ip_{2}\tau  \right) e^{ip_{2}   \tau} \sum_{\beta} e^{\beta *}_{bc} 
(\bm{p}_{2}) e^{\alpha }_{mk} (\bm{p}_{2}) \theta(-\tau)\right]
 \nonumber \\
 & \times \frac{H^{2}} {M_{\mathrm{Pl}}^{2}p_{3}^{3}} 
 \left( 1+ ip_{3}\tau'  \right) e^{-ip_{3}   \tau'} \sum_{\gamma} e^{\gamma}_{ac} 
 (\bm{p}_{3}) e^{\alpha *}_{jl} (\bm{p}_{3}) 
 \nonumber \\
& \times  \sum_{\lambda} \left( -\frac{\pi}{4}  \right) e^{-\pi \mathrm{Im}(\nu_{A}) }  
(\tau \tau')^{1/2} 
H^{(1)}_{\nu_{A}} (-k\tau) H^{(2)}_{\nu_{A}} (-k\tau') 
\varepsilon^{(\lambda)}_{0}(\bm{k}) \varepsilon^{(\lambda)}_{0}(\bm{k}) 
\nonumber \\
& \times  \sum_{\lambda'} \left( -\frac{\pi}{4}  \right) e^{-\pi \mathrm{Im}(\nu_{A}) } 
(\tau \tau')^{1/2} H^{(1)}_{\nu_{A}} (-\left|\bm{k} - \bm{p}_{1} - \bm{p}_{2} \right| \tau) 
H^{(2)}_{\nu_{A}} (-\left|\bm{k} - \bm{p}_{1} - \bm{p}_{2} \right| \tau') 
\nonumber \\
& \times \varepsilon^{(\lambda')}_{0}(\bm{k} - \bm{p}_{1} - \bm{p}_{2}) 
\varepsilon^{(\lambda')}_{0}(\bm{k} - \bm{p}_{1} - \bm{p}_{2}) 
\nonumber \\
& \times \left. k_{i}k_{j}  \left(\bm{k} - \bm{p}_{1} - \bm{p}_{2}\right)_{k} 
\left(\bm{k} - \bm{p}_{1} - \bm{p}_{2}\right)_{l} \right].
\end{align}
Introducing an expression for the sum of projections for the graviton polarization tensor as  
\begin{equation}
\label{app6}
\Pi_{ ijkl} (\bm{k}) \equiv \sum_{\alpha} e^{\alpha}_{ij} (\bm{k}) e^{\alpha *}_{kl} (\bm{k}),
\end{equation}
the equation (\ref{app5}) is written as
\begin{align}
\label{app7}
J +  \text{c.c.}  = & -2\mathrm{Re} \left[(-i)^{2} \frac{SN e^{-2\pi \mathrm{Im}(\nu_{A}) } }{36 \cdot 4^{3} \pi } 
\frac{H^{2}} {M_{\mathrm{Pl}}^{6} \left(p_{1}p_{2}p_{3} \right)^{3}} \Pi_{abim} 
(\bm{p}_{1}) \Pi_{bcmk} (\bm{p}_{2}) \Pi_{acjl} (\bm{p}_{3}) \right. 
\nonumber \\
&  \times \int_{-(1+i\epsilon)\infty}^{0} \, d\tau \int_{-(1+i\epsilon)\infty}^{0} \, d\tau'  
\frac{1}{\tau\tau'}  \left( 1- ip_{1}\tau  \right) \left( 1- ip_{2}\tau  \right) 
\left( 1+ ip_{3}\tau'  \right)  e^{i(p_{1} + p_{2})   \tau} e^{-ip_{3}   \tau'} 
\nonumber \\
& \times \int d^{3}k \,  H^{(1)}_{\nu_{A}} (-k\tau) H^{(2)}_{\nu_{A}} (-k\tau') 
H^{(1)}_{\nu_{A}} (-\left|\bm{k} - \bm{p}_{1} - \bm{p}_{2} \right| \tau) 
H^{(2)}_{\nu_{A}} (-\left|\bm{k} - \bm{p}_{1} - \bm{p}_{2} \right| \tau') 
\nonumber \\
& \times \left. k_{i}k_{j}  \left(\bm{k} - \bm{p}_{1} - \bm{p}_{2}\right)_{k}
 \left(\bm{k} - \bm{p}_{1} - \bm{p}_{2}\right)_{l} \right].
\end{align}
Approximating the Hankel functions by taking a limit $x\to\infty$ as 
\begin{equation}
\label{app8}
H^{(1)}_{\nu_{A}}(x) \to  \sqrt{ \frac{2} {\pi x} } 
e^{i \left( x - \pi/4 -  \pi \nu_{A}/2  \right)}, \quad H^{(2)}_{\nu_{A}}(x) 
\to \sqrt{ \frac{2} {\pi x} } e^{-i \left( x - \pi/4 -  \pi \nu^{*}_{A}/2  \right)}, 
\end{equation} 
 since the large momentum part is dominant in the momentum integral, 
 the momentum integral can be expressed as
\begin{align}
\label{app9}
& \int d^{3}k \,  H^{(1)}_{\nu_{A}} (-k\tau) H^{(2)}_{\nu_{A}} (-k\tau')  
H^{(1)}_{\nu_{A}} (-\left|\bm{k} - \bm{p}_{1} - \bm{p}_{2} \right| \tau) 
H^{(2)}_{\nu_{A}} (-\left|\bm{k} - \bm{p}_{1} - \bm{p}_{2} \right| \tau') 
\nonumber \\
& \times k_{i}k_{j}  \left(\bm{k} - \bm{p}_{1} - \bm{p}_{2}\right)_{k} 
\left(\bm{k} - \bm{p}_{1} - \bm{p}_{2}\right)_{l} 
\nonumber \\
= & \frac{16e^{2\pi \mathrm{Im}(\nu_{A}) }}{ \pi \tau \tau'}  
\int_{0}^{\Lambda} dk  \frac{k}{ \left|  \bm
{k} + \bm{p}_{3}\right| } e^{-i k ( \tau - \tau')} 
e^{-i \left| \bm{k} + \bm{p}_{3}\right| (\tau - \tau')}  k_{i}k_{j}  (k_{k}k_{l} + p_{3k}p_{3l})
\end{align}
where $\Lambda$ is the cutoff scale of the loop momentum 
 and the momentum integral of the odd function in $k_{i}$ is dropped.
Furthermore, by using relations for internal momentum
\begin{align}
\label{app10}
k_{i}k_{j} k_{k}k_{l} &= \frac{k^{4}}{15} \left(  \delta_{ij}  \delta_{kl}  
+ \delta_{ik}   \delta_{jl} +  \delta_{il}  \delta_{jk}   \right), \\
k_{i}k_{j} &= \frac{k^{2}}{3}  \delta_{ij} ,
\end{align}
we obtain
\begin{align}
\label{app11}
 & \frac{16e^{2\pi \mathrm{Im}(\nu_{A}) }}{ \pi \tau \tau'}  \int_{0}^{\Lambda} dk  \frac{k}{ \left|  \bm
{k} + \bm{p}_{3}\right| } e^{-i k ( \tau - \tau')} e^{-i \left|  \bm
{k} + \bm{p}_{3}\right| (\tau - \tau')}  k_{i}k_{j}  (k_{k}k_{l} + p_{3k}p_{3l}) \nonumber \\
 = & \frac{16e^{2\pi \mathrm{Im}(\nu_{A}) }}{ \pi \tau \tau'}  \int_{0}^{\Lambda} dk  \frac{k}{ \left|  \bm
{k} + \bm{p}_{3}\right| } e^{-i k ( \tau - \tau')} e^{-i \left|  \bm
{k} + \bm{p}_{3}\right| (\tau - \tau')} \nonumber \\
& \qquad \times \left[ \frac{k^{4}}{15} 
\left(  \delta_{ij} \delta_{kl}  +  \delta_{ik}   \delta_{jl} +  \delta_{il}  \delta_{jk} \right) 
+ \frac{k^{2}}{3} \delta_{ij} p_{3k}p_{3l}  \right] .
\end{align}
Assuming that the loop momentum cutoff $\Lambda$ is sufficiently larger than 
 the graviton external momentum $\bm{p}_{1,2,3}$, 
 we can approximate $\left|  \bm{k} + \bm{p}_{3}\right|$ as
\begin{equation}
\label{app12}
\left|  \bm{k} + \bm{p}_{3}\right| \simeq k\left[ 1 + \frac{p_{3} \cos \theta}{k} + \mathcal{O}
\left(\frac{p_{3}^2}{\bm{k}^2} \right) \right],
\end{equation}
where $\theta$ is the angle between $\bm{k}$ and $\bm{p}_{3}$.
From this approximation, $\tau'$-integral is performed 
\begin{align}
\label{app13}
& \int_{-(1+ i\epsilon)\infty}^{0}  d\tau' \frac{1}{\tau'^{2}}(1+ip_{3} \tau') 
e^{-i(p_{3}-k)\tau'} e^{i \left|  \bm{k} + \bm{p}_{3}\right| \tau'} + \text{c.c.} 
\nonumber \\
= & \mathrm{Re} \left[-i \int_{-(1+ i\epsilon)\infty}^{0}  d\tau' 
\frac{1}{\tau'^{2}}(1+ip_{3} \tau') e^{-i(p_{3}-k)\tau'}  e^{i  k\left( 1 + p_{3} \cos \theta / k  \right) \tau'} \right] 
\nonumber \\
= & - 2k,
\end{align}
where the formula \cite{Maldacena}
\begin{equation}
\label{app14}
\mathrm{Re} \left[-i \int_{-(1+ i\epsilon)\infty}^{0}  d\tau' \frac{1}{\tau'^{2}} (1- iK\tau)e^{-i K \tau'}\right] = -K 
\end{equation}
was utilized. 
Calculating the $\tau$-integral in $J+\text{c.c}$ in the same way, we obtain
\begin{align}
\label{app15}
 & \mathrm{Re} \left[ -i \int_{-(1+ i\epsilon)\infty}^{0} 
 d\tau \frac{1}{\tau^{2}}(1- ip_{1} \tau) (1- ip_{2} \tau) e^{-i(k-p_{1}-p_{2})\tau} 
 e^{-i \left|  \bm{k} + \bm{p}_{3}\right| \tau} \right] 
 \nonumber \\
 = & -\left( 2k -p_{1} -p_{2} - \frac{p_{1}p_{2} } {2k - p_{1} -p_{2} }  \right) .
\end{align}
Putting these together, the equation (\ref{app7}) except for the first line becomes
\begin{align}
\label{app16}
 & \frac{16e^{2\pi \mathrm{Im}(\nu_{A}) }}{ \pi } \mathrm{Re} 
 \left[(-i)^{2} \int_{-(1+i\epsilon)\infty}^{0} \, d\tau \int_{-(1+i\epsilon)\infty}^{0} \, d\tau' 
 \frac{1}{\tau\tau'}  \left( 1- ip_{1}\tau  \right) \left( 1- ip_{2}\tau  \right) 
 \left( 1+ ip_{3}\tau'  \right)  e^{i(p_{1} + p_{2}) \tau} e^{-ip_{3} \tau'} \right. 
 \nonumber \\
& \times \int d^{3}k \,  H^{(1)}_{\nu_{A}} (-k\tau) H^{(2)}_{\nu_{A}} (-k\tau') 
H^{(1)}_{\nu_{A}} (-\left|\bm{k} - \bm{p}_{1} - \bm{p}_{2} \right| \tau) 
H^{(2)}_{\nu_{A}} (-\left|\bm{k} - \bm{p}_{1} - \bm{p}_{2} \right| \tau') 
\nonumber \\
& \times \left.  k_{i}k_{j}  \left(\bm{k} - \bm{p}_{1} - \bm{p}_{2}\right)_{k} 
\left(\bm{k} - \bm{p}_{1} - \bm{p}_{2}\right)_{l} \right]  
\nonumber \\
= &\frac{16e^{2\pi \mathrm{Im}(\nu_{A}) }}{ \pi } \int_{0}^{\Lambda} dk\, 
\frac{k} {\left|  \bm{k} + \bm{p}_{3}\right| }  \left[ \frac{k^{4}}{15} 
\left(  \delta_{ij}  \delta_{kl}  +  \delta_{ik}   \delta_{jl} + \delta_{il} \delta_{jk} \right) 
+ \frac{k^{2}}{3}   \delta_{ij} p_{3k}p_{3l}  \right] \times ( - 2k ) 
\nonumber \\
& \times \left[  -\left( 2k -p_{1} -p_{2} - \frac{p_{1}p_{2} } {2k - p_{1} -p_{2} } \right)  \right] \\
\simeq &\frac{16e^{2\pi \mathrm{Im}(\nu_{A}) }}{ \pi } \int_{0}^{\Lambda} dk\, \frac{k^{4}}{15} 
\left(  \delta_{ij}  \delta_{kl}  +  \delta_{ik}   \delta_{jl} +  \delta_{il}  \delta_{jk}  \right) 
\times (-2k) \times   (-1)\left( 2k -p_{1} -p_{2}   \right) \\
\simeq &  \frac{e^{2\pi \mathrm{Im}(\nu_{A}) }}{210\pi} 
\left(  \delta_{ij} \delta_{kl} + \delta_{ik} \delta_{jl} + \delta_{il} \delta_{jk} \right) \left( 2\Lambda - p_{1} - p_{2} \right)^{7} .
\label{app17}
 \end{align}
Hence, the sum of $J$ and its complex conjugate results in
\begin{align}
J +  \text{c.c.}  = &   SN \frac{1}{36\cdot 4^{3} \pi} 
\frac{H^{2}} { M_{\mathrm{Pl}}^{6} (p_{1}p_{2}p_{3}  )^{3} } 
\Pi_{abim}(\bm{p}_{1}) \Pi_{bcmk}(\bm{p}_{2}) \Pi_{acjl}(\bm{p}_{3}) 
\nonumber \\
& \times \frac{1}{210\pi} \left(  \delta_{ij}  \delta_{kl} +  \delta_{ik}  \delta_{jl} +  \delta_{il}  \delta_{jk}   \right) 
\left(   2\Lambda - p_{1} - p_{2}  \right)^{7} \times 2.
\label{app18}
\end{align}
Note that there are still some diagrams 
 with other three-point and four-point interactions in (\ref{3pt}) and (\ref{4pt}), 
 which we should calculate in the first category. 
We can easily check that the contributions from these diagrams are roughly same order as the result of (\ref{app18}). 

Next, we discuss the other diagram from the latter category, 
which will be denoted by $J_m$. 
As in the previous calculations of $J$, 
we calculate a sum of $J_m$ and its complex conjugate. 
\begin{align}
\label{app19}
J_m + \text{c.c.} =& 2 \text{Re} 
\left[ SN   \int_{-(1+i\epsilon)\infty}^{0}a(\tau) \, d\tau \int_{-(1+i\epsilon)\infty}^{0}a(\tau') 
\, d\tau' \int \frac{d^{3}k} {(2\pi)^{3}} i \cdot (-i) \frac{1}{2} a(\tau)  \left(  -a(\tau')  \right) m_X^4
\right. \nonumber \\
& \times  G_{++}^{ab, im}(\tau , 0, \bm{p}_{1}) G_{++}^{bc, mk}(\tau , 0, \bm{p}_{2}) G_{-+}^{ac, jl}(\tau' , 0, \bm{p}_{3}) 
\nonumber \\
& \times \frac{1}{3} \sum_{\lambda} \Delta_{ij +-}^{(\lambda)} ( \tau, \tau', k) 
\times \left. \frac{1}{3} \sum_{\lambda'} \Delta_{kl +-}^{(\lambda')} 
\left( \tau, \tau', \left| \bm{k} - \bm{p}_{1} - \bm{p}_{2} \right| \right) \right].
\end{align}
Substituting the graviton propagator (\ref{app2}) 
 and the $X, Y$ gauge boson propagators (\ref{app3}), (\ref{app4}) into (\ref{app19}), 
we obtain
\begin{align}
J_m + \text{c.c.} =& 2 \text{Re} \left[ (-i)^2 \frac{SN}{36} m_X^4 
\int_{-(1+i\epsilon)\infty}^{0} \, d\tau \int_{-(1+i\epsilon)\infty}^{0} \, d\tau' 
\int \frac{d^{3}k} {(2\pi)^{3}} \left( -\frac{1}{H\tau}  \right)^{2} \left( -\frac{1}{H\tau'}  \right)^{2} 
\right.  \nonumber \\
& \times \left[ \frac{H^{2}} {M_{\mathrm{Pl}}^{2}p_{1}^{3}}  
\left( 1+ ip_{1}\tau  \right) e^{-ip_{1}   \tau} 
\sum_{\alpha} e^{\alpha}_{ab} (\bm{p}_{1}) e^{\alpha *}_{im} (\bm{p}_{1}) \theta(\tau) \right. 
\nonumber \\
 & \left. \quad + \frac{H^{2}} {M_{\mathrm{Pl}}^{2}p_{1}^{3}} 
 \left( 1- ip_{1}\tau  \right) e^{ip_{1}   \tau} \sum_{\alpha} e^{\alpha *}_{ab} (\bm{p}_{1}) 
 e^{\alpha }_{im} (\bm{p}_{1}) \theta(-\tau)\right]  
 \nonumber \\
& \times \left[ \frac{H^{2}} {M_{\mathrm{Pl}}^{2}p_{2}^{3}} 
\left( 1+ ip_{2}\tau  \right) e^{-ip_{2} \tau} \sum_{\beta} 
e^{\beta}_{bc} (\bm{p}_{2}) e^{\alpha *}_{mk} (\bm{p}_{2}) \theta(\tau) \right. 
\nonumber \\
& \left. \quad + \frac{H^{2}} {M_{\mathrm{Pl}}^{2}p_{2}^{3}} 
\left( 1- ip_{2}\tau  \right) e^{ip_{2}   \tau} \sum_{\beta} e^{\beta *}_{bc} 
(\bm{p}_{2}) e^{\alpha }_{mk} (\bm{p}_{2}) \theta(-\tau)\right]
 \nonumber \\
 & \times \frac{H^{2}} {M_{\mathrm{Pl}}^{2}p_{3}^{3}} 
 \left( 1+ ip_{3}\tau'  \right) e^{-ip_{3}   \tau'} \sum_{\gamma} e^{\gamma}_{ac} 
 (\bm{p}_{3}) e^{\alpha *}_{jl} (\bm{p}_{3}) 
 \nonumber \\
& \times  \sum_{\lambda} \left( -\frac{\pi}{4}  \right) e^{-\pi \mathrm{Im}(\nu_{A}) }  
(\tau \tau')^{1/2} H^{(1)}_{\nu_{A}} (-k\tau) H^{(2)}_{\nu_{A}} (-k\tau') 
\varepsilon^{(\lambda)}_{i}(\bm{k}) \varepsilon^{(\lambda)}_{j}(\bm{k}) 
\nonumber \\
& \times  \sum_{\lambda'} \left( -\frac{\pi}{4}  \right) e^{-\pi \mathrm{Im}(\nu_{A}) } 
(\tau \tau')^{1/2} H^{(1)}_{\nu_{A}} (-\left|\bm{k} - \bm{p}_{1} - \bm{p}_{2} \right| \tau) 
H^{(2)}_{\nu_{A}} (-\left|\bm{k} - \bm{p}_{1} - \bm{p}_{2} \right| \tau') 
\nonumber \\
& \times \left. \varepsilon^{(\lambda')}_{k}(\bm{k} - \bm{p}_{1} - \bm{p}_{2}) 
\varepsilon^{(\lambda')}_{l}(\bm{k} - \bm{p}_{1} - \bm{p}_{2}) \right]. 
\end{align}
By using the approximation of Hankel functions and the quantity $\Pi_{abcd}$, 
we can further rewrite the above expression as
\begin{align}
J_m + \text{c.c.} \simeq & 2 \text{Re} \left[  
-SN \frac{1}{32 \cdot 3^{2} \pi^3 } 
\frac{H^{2} m_X^4}{M_{\mathrm{Pl}}^{6} \left(p_{1}p_{2}p_{3} \right)^{3}} 
\Pi_{abim} (\bm{p}_{1}) \Pi_{bcmk} (\bm{p}_{2}) \Pi_{acjl} (\bm{p}_{3}) \right.
\nonumber \\
&  \times \int_{-(1+i\epsilon)\infty}^{0} \, d\tau \int_{-(1+i\epsilon)\infty}^{0} \, d\tau'  
\frac{1}{(\tau\tau')^2}  \left( 1- ip_{1}\tau  \right) \left( 1- ip_{2}\tau  \right) 
\left( 1+ ip_{3}\tau'  \right) 
\nonumber \\
& \times \int d^{3}k \frac{1}{k^2} 
\sum_{\lambda, \lambda'} 
\varepsilon_i^{(\lambda)}(\bm{k})
\varepsilon_j^{(\lambda)}(\bm{k})
\varepsilon_k^{(\lambda')}(\bm{k} - \bm{p}_{1} - \bm{p}_{2})
\varepsilon_l^{(\lambda')}(\bm{k} - \bm{p}_{1} - \bm{p}_{2}) 
\nonumber \\
& \times \left. 
e^{-i(k-p_1-p_2)\tau} e^{-i|\bm{k} - \bm{p}_{1} - \bm{p}_{2}|\tau} 
e^{-i(-k+p_3) \tau'} e^{-i| \bm{k} - \bm{p}_{1} - \bm{p}_{2}| \tau'} \right]
\nonumber \\
\simeq & 2 \text{Re} \left[ 
-SN \frac{1}{8 \cdot 3^{2} \pi^2 } 
\frac{H^{2} m_X^4}{M_{\mathrm{Pl}}^{6} \left(p_{1}p_{2}p_{3} \right)^{3}} 
\Pi_{abim} (\bm{p}_{1}) \Pi_{bcmk} (\bm{p}_{2}) \Pi_{acik} (\bm{p}_{3}) \right. 
\nonumber \\
&  \times \int_{-(1+i\epsilon)\infty}^{0} \, d\tau \int_{-(1+i\epsilon)\infty}^{0} \, d\tau'  
\frac{1}{(\tau\tau')^2}  \left( 1- ip_{1}\tau  \right) \left( 1- ip_{2}\tau  \right) 
\left( 1+ ip_{3}\tau'  \right) 
\nonumber \\
& \times \left. \int_0^\Lambda dk \frac{1}{(H \tau)^4} 
e^{-i(k-p_1-p_2)\tau} e^{-i(k + p_3 \cos \theta) \tau} 
e^{-i(-k+p_3) \tau'} e^{-i(k + p_3 \cos \theta) \tau'} \right]
\nonumber \\
=& J \times 2^4 \times 30 \times \left( \frac{m_X}{H} \right)^4. 
\label{app20}
\end{align}
In the second approximation, 
 the sum of the product of polarization vectors for graviton is approximated as
\begin{align}
\sum_{\lambda} 
\varepsilon_i^{(\lambda)}(\bm{k})
\varepsilon_j^{(\lambda)}(\bm{k})
= a(\tau)^2\delta_{ij} + \frac{k_i k_j}{m_X^2} 
= \frac{\delta_{ij}}{(H\tau)^2} + \frac{k_i k_j}{m_X^2} 
\simeq \frac{\delta_{ij}}{(H \tau)^2}
\label{app21}
\end{align}
by taking into account of the assumption $H^2 \ll m_X^2$. 
Assumption that the cutoff $\Lambda$ is sufficiently larger than 
 the graviton external momentum $\bm{p}_{1,2,3}$ is also understood in the momentum integral. 



\begin{thebibliography}{100}
\bibitem{ChenWang1} 
X. Chen and Y. Wang, ``Large non-Gaussianities with Intermediate Shapes from Quasi- Single Field Inflation," 
{\it Phys. Rev.} \textbf{D 81}, 063511 (2010) [arXiv:0909.0496 [astro-ph.CO]].

\bibitem{ChenWang2}
X. Chen and Y. Wang, ``Quasi-Single Field Inflation and Non-Gaussianities," 
{\it JCAP} \textbf{04}, 027 (2010) [arXiv:0911.3380 [hep-th]].

\bibitem{BaumannGreen}
D. Baumann and D. Green, ``Signatures of Supersymmetry from the Early Universe,"  
{\it Phys. Rev.} \textbf{D 85}, 103520 (2012) [arXiv:1109.0292 [hep-th]].

\bibitem{ABG}
V. Assassi, D. Baumann and D. Green, ``On Soft Limits of Inflationary Correlation Functions,"  
{\it JCAP} \textbf{11}, 047 (2012) [arXiv:1204.4207 [hep-th]].

\bibitem{SFCS}
E. Sefusatti, J. R. Fergusson, X. Chen and E. P. S. Shellard, 
``Effects and Detectability of Quasi-Single Field Inflation in the Large-Scale Structure and Cosmic Microwave Background,"  
{\it JCAP} \textbf{08}, 033 (2012) [arXiv:1204.6318 [astro-ph.CO]].

\bibitem{NVBB}
J. Norena, L. Verde, G. Barenboim and C. Bosch, 
``Prospects for constraining the shape of non-Gaussianity with the scale-dependent bias,"  
{\it JCAP} \textbf{08}, 019 (2012) [arXiv:1204.6324 [astro-ph.CO]].
   
\bibitem{ChenWang3}
X. Chen and Y. Wang, ``Quasi-Single Field Inflation with Large Mass,"  
{\it JCAP} \textbf{09}, 021 (2012) [arXiv:1205.0160 [hep-th]].

\bibitem{NYY}
T. Noumi, M. Yamaguchi and D. Yokoyama, 
``Effective field theory approach to quasi-single field inflation and effects of heavy fields,"  
{\it JHEP} \textbf{06}, 051 (2013) [arXiv:1211.1624 [hep-th]].

\bibitem{GPS}
J. O. Gong, S. Pi and M. Sasaki, ``Equilateral non-Gaussianity from heavy fields,"  
{\it JCAP} \textbf{11}, 043 (2013) [arXiv:1306.3691 [hep-th]].

\bibitem{Emami}
R. Emami, ``Spectroscopy of Masses and Couplings during Inflation,"  
{\it JCAP} \textbf{04}, 031 (2014) [arXiv:1311.0184 [hep-th]].

\bibitem{KehagiasRiotto1}
A. Kehagias and A. Riotto, ``High Energy Physics Signatures from Inflation and Con-formal Symmetry of de Sitter,"  
{\it Fortsch. Phys.} \textbf{63}, 531 (2015) [arXiv:1501.03515 [hep-th]].

\bibitem{LWZ}
J. Liu, Y. Wang and S. Zhou, ``Inflation with Massive Vector Fields,"  
{\it JCAP} \textbf{08}, 033 (2015) [arXiv:1502.05138 [hep-th]].

\bibitem{NAHM}
N. Arkani-Hamed and J. Maldacena, 
``Cosmological Collider Physics," arXiv:1503.08043 [hep-th].

\bibitem{DFK}
E. Dimastrogiovanni, M. Fasiello and M. Kamionkowski, 
``Imprints of Massive Primordial Fields on Large-Scale Structure," 
{\it JCAP} \textbf{02}, 017 (2016) [arXiv:1504.05993 [astro- ph.CO]].

\bibitem{SCD}
F. Schmidt, N. E. Chisari and C. Dvorkin, ``Imprint of inflation on galaxy shape correlations," 
{\it JCAP} \textbf{10}, 032 (2015) [arXiv:1506.02671 [astro-ph.CO]].

\bibitem{CNW}
X. Chen, M. H. Namjoo and Y. Wang, ``Quantum Primordial Standard Clocks," 
{\it JCAP} \textbf{02}, 013 (2016) [arXiv:1509.03930 [astro-ph.CO]].

\bibitem{DNS}
L. V. Delacretaz, T. Noumi and L. Senatore, ``Boost Breaking in the EFT of Inflation," 
{\it JCAP} \textbf{02}, 034 (2017) [arXiv:1512.04100 [hep-th]].

\bibitem{BBDS}
B. Bonga, S. Brahma, A. S. Deutsch and S. Shandera, ``Cosmic variance in inflation with two light scalars," 
{\it JCAP} \textbf{05}, 018 (2016) [arXiv:1512.05365 [astro-ph.CO]].

\bibitem{FMSS}
R. Flauger, M. Mirbabayi, L. Senatore and E. Silverstein, 
``Productive Interactions: heavy particles and non-Gaussianity," 
{\it JCAP} \textbf{10}, 058 (2017) [arXiv:1606.00513 [hep- th]].

\bibitem{LBP}
H. Lee, D. Baumann and G. L. Pimentel, ``Non-Gaussianity as a Particle Detector," 
{\it JHEP} \textbf{12}, 040 (2016) [arXiv:1607.03735 [hep-th]].

\bibitem{DGS}
L. V. Delacretaz, V. Gorbenko and L. Senatore, ``The Supersymmetric Effective Field Theory of Inflation," 
{\it JHEP} \textbf{03}, 063 (2017) [arXiv:1610.04227 [hep-th]].

\bibitem{MMMC}
P. D. Meerburg, M. M\"unchmeyer, J. B. Mu\~noz and X. Chen, ``Prospects for Cosmological Collider Physics,"
{\it JCAP} \textbf{03}, 050 (2017) [arXiv:1610.06559 [astro-ph.CO]].

\bibitem{CWX1}
X. Chen, Y. Wang and Z. Z. Xianyu, ``Standard Model Background of the Cosmological Collider," 
{\it Phys. Rev. Lett.} \textbf{118}, no.26, 261302 (2017) [arXiv:1610.06597 [hep-th]].

\bibitem{CWX2}
X. Chen, Y. Wang and Z. Z. Xianyu, ``Standard Model Mass Spectrum in Inflationary Universe," 
{\it JHEP} \textbf{04}, 058 (2017) [arXiv:1612.08122 [hep-th]].

\bibitem{KehagiasRiotto2}
A. Kehagias and A. Riotto, ``On the Inflationary Perturbations of Massive Higher-Spin Fields," 
{\it JCAP} \textbf{07}, 046 (2017) [arXiv:1705.05834 [hep-th]].

\bibitem{AMRW1}
H. An, M. McAneny, A. K. Ridgway and M. B. Wise, ``Quasi Single Field Inflation in the non-perturbative regime," 
{\it JHEP} \textbf{06}, 105 (2018) [arXiv:1706.09971 [hep-ph]].

\bibitem{TWZ}
X. Tong, Y. Wang and S. Zhou, ``On the Effective Field Theory for Quasi-Single Field Inflation," 
{\it JCAP} \textbf{11}, 045 (2017) [arXiv:1708.01709 [astro-ph.CO]].

\bibitem{IWWZ}
A. V. Iyer, S. Pi, Y. Wang, Z. Wang and S. Zhou, ``Strongly Coupled Quasi-Single Field Inflation," 
{\it JCAP} \textbf{01}, 041 (2018) [arXiv:1710.03054 [hep-th]].

\bibitem{AMRW2}
H. An, M. McAneny, A. K. Ridgway and M. B. Wise, 
``Non-Gaussian Enhancements of Galactic Halo Correlations in Quasi-Single Field Inflation," 
{\it Phys. Rev.} \textbf{D97}, no.12, 123528 (2018) [arXiv:1711.02667 [hep-ph]].

\bibitem{KumarSundrum1}
S. Kumar and R. Sundrum, ``Heavy-Lifting of Gauge Theories By Cosmic Inflation," 
{\it JHEP} \textbf{05}, 011 (2018) [arXiv:1711.03988 [hep-ph]].

\bibitem{Riquelme}
S. Riquelme M., ``Non-Gaussianities in a two-field generalization of Natural Inflation," 
{\it JCAP} \textbf{04}, 027 (2018) [arXiv:1711.08549 [astro-ph.CO]].

\bibitem{FKR}
G. Franciolini, A. Kehagias and A. Riotto, ``Imprints of Spinning Particles on Primordial Cosmological Perturbations," 
{\it JCAP} \textbf{02}, 023 (2018) [arXiv:1712.06626 [hep-th]].

\bibitem{SaitoKubota}
R. Saito and T. Kubota, ``Heavy Particle Signatures in Cosmological Correlation Functions with Tensor Modes," 
{\it JCAP} \textbf{06}, 009 (2018) [arXiv:1804.06974 [hep-th]].

\bibitem{CPS}
G. Cabass, E. Pajer and F. Schmidt, ``Imprints of Oscillatory Bispectra on Galaxy Clustering," 
{\it JCAP} \textbf{09}, 003 (2018) [arXiv:1804.07295 [astro-ph.CO]].

\bibitem{WWYZ}
Y. Wang, Y. P. Wu, J. Yokoyama and S. Zhou, ``Hybrid Quasi-Single Field Inflation," 
{\it JCAP} \textbf{07}, 068 (2018) [arXiv:1804.07541 [astro-ph.CO]]. 

\bibitem{CWX3}
X. Chen, Y. Wang and Z. Z. Xianyu, ``Neutrino Signatures in Primordial Non-Gaussianities," 
{\it JHEP} \textbf{09}, 022 (2018) [arXiv:1805.02656 [hep-ph]].

\bibitem{DFT}
E. Dimastrogiovanni, M. Fasiello and G. Tasinato, ``Probing the inflationary particle content: extra spin-2 field," 
{\it JCAP} \textbf{08}, 016 (2018) [arXiv:1806.00850 [astro-ph.CO]].

\bibitem{BCKS}
L. Bordin, P. Creminelli, A. Khmelnitsky and L. Senatore, ``Light Particles with Spin in Inflation," 
{\it JCAP} \textbf{10}, 013 (2018) [arXiv:1806.10587 [hep-th]].

\bibitem{CDWZ}
W. Z. Chua, Q. Ding, Y. Wang and S. Zhou, ``Imprints of Schwinger Effect on Primordial Spectra," 
{\it JHEP} \textbf{04}, 066 (2019) [arXiv:1810.09815 [hep-th]].

\bibitem{NAHBLP}
N. Arkani-Hamed, D. Baumann, H. Lee and G. L. Pimentel, 
``The Cosmological Bootstrap: Inflationary Correlators from Symmetries and Singularities," 
{\it JHEP} \textbf{04}, 105 (2020) [arXiv:1811.00024 [hep-th]].

\bibitem{KumarSundrum2}
S. Kumar and R. Sundrum, ``Seeing Higher-Dimensional Grand Unification In Primordial Non-Gaussianities," 
{\it JHEP} \textbf{04}, 120 (2019) [arXiv:1811.11200 [hep-ph]].

\bibitem{GHJT}
G. Goon, K. Hinterbichler, A. Joyce and M. Trodden,  
``Shapes of gravity: Tensor non-Gaussianity and massive spin-2 fields," 
{\it JHEP} \textbf{10}, 182 (2019) [arXiv:1812.07571 [hep-th]].

\bibitem{Wu}
Y. P. Wu, ``Higgs as heavy-lifted physics during inflation," 
{\it JHEP} \textbf{04}, 125 (2019) [arXiv:1812.10654 [hep-ph]].

\bibitem{ADLFKR}
D. Anninos, V. De Luca, G. Franciolini, A. Kehagias and A. Riotto, ``Cosmological Shapes of Higher-Spin Gravity," 
{\it JCAP} \textbf{04}, 045 (2019) [arXiv:1902.01251 [hep-th]].

\bibitem{NSWZ}
L. Li, T. Nakama, C. M. Sou, Y. Wang and S. Zhou, 
``Gravitational Production of Superheavy Dark Matter and Associated Cosmological Signatures," 
{\it JHEP} \textbf{07}, 067 (2019) [arXiv:1903.08842 [astro-ph.CO]].

\bibitem{MR}
M. McAneny and A. K. Ridgway, 
``New Shapes of Primordial Non-Gaussianity from Quasi-Single Field Inflation with Multiple Isocurvatons," 
{\it Phys. Rev.} \textbf{D100}, no.4, 043534 (2019) [arXiv:1903.11607 [astro-ph.CO]].

\bibitem{KNTZ}
S. Kim, T. Noumi, K. Takeuchi and S. Zhou, 
``Heavy Spinning Particles from Signs of Primordial Non-Gaussianities: Beyond the Positivity Bounds," 
{\it JHEP} \textbf{12}, 107 (2019) [arXiv:1906.11840 [hep-th]].

\bibitem{LWX}
S. Lu, Y. Wang and Z. Z. Xianyu, ``A Cosmological Higgs Collider," 
{\it JHEP} \textbf{02}, 011 (2020) [arXiv:1907.07390 [hep-th]].

\bibitem{HookHaungRacco1}
A. Hook, J. Huang and D. Racco, ``Searches for other vacua. Part II. A new Higgstory at the cosmological collider," 
{\it JHEP} \textbf{01}, 105 (2020) [arXiv:1907.10624 [hep-ph]].
25

\bibitem{HookHuangRacco2}
A. Hook, J. Huang and D. Racco, ``Minimal signatures of the Standard Model in non-Gaussianities," 
{\it Phys. Rev.} \textbf{D 101}, no.2, 023519 (2020) [arXiv:1908.00019 [hep-ph]].

\bibitem{KumarSundrum3}
S. Kumar and R. Sundrum, ``Cosmological Collider Physics and the Curvaton," 
{\it JHEP} \textbf{04}, 077 (2020) [arXiv:1908.11378 [hep-ph]].

\bibitem{WangXianyu1}
L. T. Wang and Z. Z. Xianyu, ``In Search of Large Signals at the Cosmological Collider," 
{\it JHEP} \textbf{02}, 044 (2020) [arXiv:1910.12876 [hep-ph]].

\bibitem{WangZhu}
Y. Wang and Y. Zhu, ``Cosmological Collider Signatures of Massive Vectors from Non-Gaussian Gravitational Waves," 
{\it JCAP} \textbf{04}, 049 (2020) [arXiv:2001.03879 [astro-ph.CO]].

\bibitem{LLWZ}
L. Li, S. Lu, Y. Wang and S. Zhou, ``Cosmological Signatures of Superheavy Dark Matter," 
{\it JHEP} \textbf{07}, 231 (2020) [arXiv:2002.01131 [hep-ph]].

\bibitem{WangXianyu2}
L. T. Wang and Z. Z. Xianyu, ``Gauge Boson Signals at the Cosmological Collider," 
{\it JHEP} \textbf{11}, 082 (2020) [arXiv:2004.02887 [hep-ph]].

\bibitem{BCS}
A. Bodas, S. Kumar and R. Sundrum, ``The Scalar Chemical Potential in Cosmological Collider Physics," 
[arXiv:2010.04727 [hep-ph]].




\bibitem{Maldacena} 
J. Maldacena,  ``Non-Gaussian features of primordial fluctuations in single field inflationary models,'' 
 {\it JHEP} \textbf{0305}, 013(2003), arXiv:0210603\ [astro-ph].

\bibitem{Cheung}
C. Cheung, {\it et al}., ``The Effective Field Theory of Inflation," 
   {\it JHEP} \textbf{0803}, 014(2008), arXiv:0709.0293\ [hep-th].

\bibitem{Leonardo} 
L. Senatore, ``Lectures on Inflation,'' arXiv:1609.00716\ [hep-th].

\bibitem{Ryo Saito}  
R. Saito,  ``Cosmological correlation functions including a massive scalar field and an arbitrary number of soft-gravitons,'' 
   arXiv:1803.01287\ [hep-th].

\bibitem{SK} X. Chen, Y. Wang, and Z. Z. Xianyu, 
 ``Schwinger-Keldysh Diagrammatics for Primordial Perturbations,'' 
  {\it JCAP} \textbf{1712}, 006(2017), arXiv:1703.10166\ [hep-th].

\bibitem{Chen}
X. Chen, ``Primordial Non-Gaussianities from Inflation Models,'' 
  {\it Adv.Astron} 2010 (2010) 638979, arXiv:1002.1416\ [astro-ph.CO].

\bibitem{Baumann2018} 
D. Baumann, ``TASI Lectures on Primordial Cosmology,'' 
 arXiv:1807.03098\ [hep-th].

\bibitem{in-in}
J. Schwinger, ``Brownian Motion of a Quantum Oscillator,” {\it J. Math. Phys.} \textbf{2} (1961) 407; 
R. Jordan, ``Effective Field Equations for Expectation Values,” 
{\it Phys. Rev.} \textbf{D33} (1986) 444; 
E. Calzetta and B. Hu, 
 ``Closed Time Path Functional Formalism in Curved Space-Time: Application to Cosmological Back Reaction Problems,” 
 {\it Phys. Rev.} \textbf{D35} (1987) 495;
J. Maldacena, ``Non-Gaussian Features of Primordial Fluctuations in Single-Field Inflationary Models,” 
 JHEP 05 (2003) 013, arXiv: astro-ph/0210603 [astro-ph]; 
S. Weinberg, ``Quantum Contributions to Cosmological Correlations,” 
{\it Phys. Rev.} \textbf{D72} (2005) 043514, arXiv: hep-th/0506236 [hep-th]. 



\end{thebibliography}
\end{document}